\documentclass[12pt,english]{article}
\usepackage[T1]{fontenc}
\usepackage[latin9]{inputenc}
\usepackage[a4paper]{geometry}
\usepackage{xcolor}
\usepackage{float}
\usepackage{graphicx}

\makeatletter
\newcommand{\lyxaddress}[1]{
	\par {\raggedright #1
	\vspace{1.4em}
	\noindent\par}
}

\usepackage{babel}

\makeatother

\usepackage{babel}
\begin{document}

\title{Control of noise in gene expression by transcriptional reinitiation}

\author{Rajesh Karmakar\thanks{Electronic address:rkarmakar@vidyamandira.ac.in}}
\maketitle

\lyxaddress{}

\lyxaddress{\begin{center}
Ramakrishna Mission Vidyamandira, Belur Math, Belur, Howrah, West
Bengal, India 
\par\end{center}}
\begin{abstract}
Gene expression is a random or noisy process. The process consists
of several random events among which the reinitiation of transcription
by RNAP is an important one. The RNAP molecules can bind the gene
only after the promoter gets activated by transcription factors. Several
transcription factors bind the promoter to put the gene in the active
state. The gene turns into inactive state as the bound transcription
factors leave the promoter. During the active period of the gene,
many RNAP molecules transcribe the gene to synthesize the mRNAs. The
binding event of RNAP to the active state of the gene is a probabilistic
process and therefore, introduces noise or fluctuations in the mRNA
and protein levels. In this paper, we analytically calculate the Fano
factor in mRNA and protein levels and also the probability distribution
of mRNA numbers exactly with the binding event of RNAPs in gene transcription
process. The analytically calculated expression of Fano factor of
proteins shows excellent agreement with an experimental result. Then
we show that the Fano factor in mRNA levels can be sub-Poissonian
due to the reinitiation of transcription by RNAP and the mean mRNA
level can be increased without increasing the Fano factor. Our study
show that the Fano factor can also be reduced keeping mRNA levels
fixed. We find that the reinitiation of transcription can behave as
a fine-tuned control process to regulate the mRNA/protein level in
the cell.

Key words: Stochastic gene expression, Reinitiation of transcription,
Fano factor 
\end{abstract}

\section{Introduction}

Gene expression and its regulation is an inherently stochastic or
noisy process and this stochasticity produces cell-to-cell variation
in mRNA and protein levels. The variation or fluctuations in mRNA
and protein levels is qualitatively known as noise. The genetically
identical individuals can be very different even if they grow in an
identical environmental condition because of noisy gene expression.
There are large number of theoretical and experimental studies at
the single cell level which establish the stochastic nature of gene
expression \cite{key-109,key-110,key-110A,key-111,key-111A,key-112,key-113,key-114,key-115,key-116,key-117,key-117B,key-118,key-119,key-121,key-121A}.
The effect of noise or fluctuations in gene expression can be beneficial
or detrimental to the progression of cellular activities \cite{key-111A,key-115,key-122A,key-122B,key-122C,key-122D}.
Variability in mRNA and protein levels can be beneficial to adapt
\textbf{\textcolor{brown}{to}} sudden changes in environmental conditions
\cite{key-122A}. The noise in gene expression can also generate phenotypic
diversity in genetically identical cells and may play an important
role in the survival of cell colonies \cite{key-111A}. Fluctuations
in mRNA and protein levels can also have detrimental role in functioning
the fine-tuned cellular processes \cite{key-122C}. Different studies
show that the cellular system has its own mechanism to reduce noise
or fluctuations in gene expression. There is a study showing that
noise levels in proteins from essential genes are lower than that
from nonessential genes \cite{key-122E}. It has also been shown that
organisms could evolve to minimize the variability in gene expression
by adopting special technique e.g., high transcription rate and low
translation rate. The noise strength does not remain fixed throughout
the life of a cell rather it decreases during the normal aging of
a cell \cite{key-122F}. Many other processes exist by which the noise
in gene expression can be reduced \cite{key-13,key-14}.

The gene expression and regulation consists of several important biochemical
steps e.g., transcription, translation, degradation etc. The regulation
can be achieved at the transcription or translation or degradation
level and the control of transcription is the dominant form of regulation
of gene expression. The transcriptional regulation is mostly done
by DNA binding proteins known as transcription factors. A group of
transcription factors (TFIIA, TBP and associated factors) bind the
promoter sequences on the DNA and turn the gene into the active state
or intermediate promoter complex \cite{key-3,key-6}. The transcription
factors which modulate the transitions to the active (inactive) state
are known as activators (repressors). Gene can either be in the active
or inactive states in the cell and random transitions take place between
the two states \cite{key-112,key-113,key-114,key-115,key-116,key-117}.
In the active state, the RNA polymerase (RNAP) proteins bind the gene
and form a preinitiation complex (including RNAP). The bound RNAP
then initiate the transcription of the protein-coding gene to synthesize
mRNAs \cite{key-111,key-111A}. The RNAP molecules transcribe the
gene several times and synthesize a burst of mRNA. That is known as
reinitiation of transcription by RNAP. The initiation of transcription
by RNAP molecules is an important part of the transcriptional regulation
\cite{key-123}. The control of transcription at the initiation stage
is a key mechanism for the regulation of gene expression.

There are some gene regulatory systems where activators and repressors
are both involved in transcriptional regulation \cite{key-111,key-111A,key-117B,key-118}.
It has been shown that the gene can be in more than two states under
the regulation of activators and repressors \cite{key-111,key-111A,key-118,key-124A}.
It has also been shown that the stochastic effects due to random transitions
between active and inactive states of a gene are much stronger than
the stochastic effects caused by random production and degradation
of single mRNA and protein molecules \cite{key-111,key-111A,key-113,key-115,key-116,key-124A,key-124B}.
In the course of time, different model systems have been developed
to explain the experimentally observed variability in gene expression.
The simplest model is the constitutive gene expression in which gene
is always in the active state and mRNAs are synthesized at constant
rate from that state \cite{key-110,key-117,key-124B,key-124C}. In
such case, the fluctuations in mRNA level arise from the random birth
and death of mRNAs. The study of cell-to-cell variation in mRNA and
protein levels at the single cell level is mostly done by assuming
two states of a gene under the regulation of activators \cite{key-112,key-113,key-114,key-115,key-116,key-117,key-119,key-121}.
In the two-state model, the randomness in mRNA levels comes from the
random transitions between the gene states and from the random birth
and death of mRNAs from the active gene \cite{key-121,key-124B,key-124C}.
But, the stochasticity due to reinitiation of transcription by RNAP
is ignored in both constitutive and two-state gene activation models.
Blake et al. experimentally study the synthetic GAL1{*} yeast promoter
and identify the regulatory mechanism using stochastic simulations
that agrees well with their experimental data. They observed that
the pulsatile mRNA production through reinitiation is crucial for
the dependence of noise or fluctuations on transcriptional efficiency
\cite{key-111,key-111A}. It has been shown that the reinitiation
also enhances the reliability of transcriptional \textbf{\textcolor{black}{response}}
in eukaryotic systems \cite{key-124}. Sanchez et al. studied the
GAL1{*} yeast promoter also and analytically calculate the various
moments neglecting the reinitiation step in gene transcription \cite{key-124A}.
Though it has been identified that the reinitiation of transcription
by RNAP plays crucial role in the phenotypic variability in cells
but its functional role in noisy mRNA or protein levels is very poorly
understood.

The stochasticity in gene expression is generally quantified by calculating
the variance or coefficient of variation (CV) or Fano factor \cite{key-110,key-110A,key-111,key-113,key-121,key-124A,key-124B,key-124C}.
The mRNA synthesis from a gene constitutively is a Poisson process
with unity Fano factor ($FF$). The two-state model gives rise to
an increased FF in mRNA level ($FF_{m}$) due to random transitions
between gene states. The $FF_{m}$ now becomes super-Poissonian (greater
than unity) \cite{key-121,key-124B}. It is observed that only negative
feedback can reduce the $FF_{m}$ and can move that into the sub-Poissonian
(less than unity) regime \cite{key-119}. With the two-state gene
activation model, we cannot think of the sub-Poissonian Fano factor
regime at the mRNA level without the negative feedback in gene expression.
In this article, we show that the reinitiation of transcription by
RNAP in two-state gene activation process can move the $FF_{m}$ into
the sub-Poissonian regime. We observe that the mean mRNA level can
be increased without increasing the $FF_{m}$ by controlling the reinitiation
of transcription. We also observe that $FF_{m}$ can be decreased
to sub-Poissonian regime keeping average mRNA level fixed. Again,
the knowledge of FF may not always give the complete information of
the proteins. The distribution of mRNA/protein level helps to identify
the more important information about responses of gene expression.
Obtaining an exact analytical expression for the distribution of mRNA/protein
has been a challenging task and that always gives some extra insights.
There are many works that have already been done on the exact solution
of distribution of mRNA and protein levels with and without the feedback
but excluding the reinitiation process \cite{key-114,key-117,key-118,key-119}.
We also find out the exact steady state distribution of mRNA levels
with reinitiation of transcription by RNAP.

\section{The model and analysis}

\subsection{The model}

We consider the gene regulation model where only activators regulate
the gene transcription. In this model, the activators bind the promoter
region of the gene and turn the gene into active state ($G_{2}$)
from the inactive state ($G_{1}$). Once the gene turns into active
state, the RNAP then binds the gene and form an initiation complex
($G_{3}$). Then, bound RNAP can follow two paths: either it unbinds
and leaves the $G_{3}$ state or it starts transcription. As it starts
transcription, it moves forward along the gene and the $G_{3}$ state
turns into the $G_{2}$ again. Another RNAP can bind the gene again
to form an initiation complex and then starts transcription. So, RNAP
can bind multiple times in the active state of the gene and a burst
of mRNAs are synthesized before the gene turns into the inactive state
($G_{1}$) from the active state ($G_{2}$). The biochemical reactions
and the corresponding rate constants are shown in figure 1. The model
considered here is a part of the network identified in synthetic GAL1{*}
promoter by Blake et al. \cite{key-111}. The GAL1{*} promoter is
regulated by activator and repressor both and their concentrations
are further controlled by external inducer GAL and ATC respectively.
In experiment, the activation (repression) probability of the promoter
increases (decreases) with the increase of GAL (ATC) concentrations
\cite{key-111}. The GAL1{*} promoter may behave as an activator-only
system with the full induction of ATC ({[}ATC{]} = 500ng/ml). Thus,
the promoter architecture proposed by them has the resemblance with
our model given in figure 1 with full ATC induction.

\begin{figure}[H]
\begin{centering}
\includegraphics[width=6cm,height=2cm]{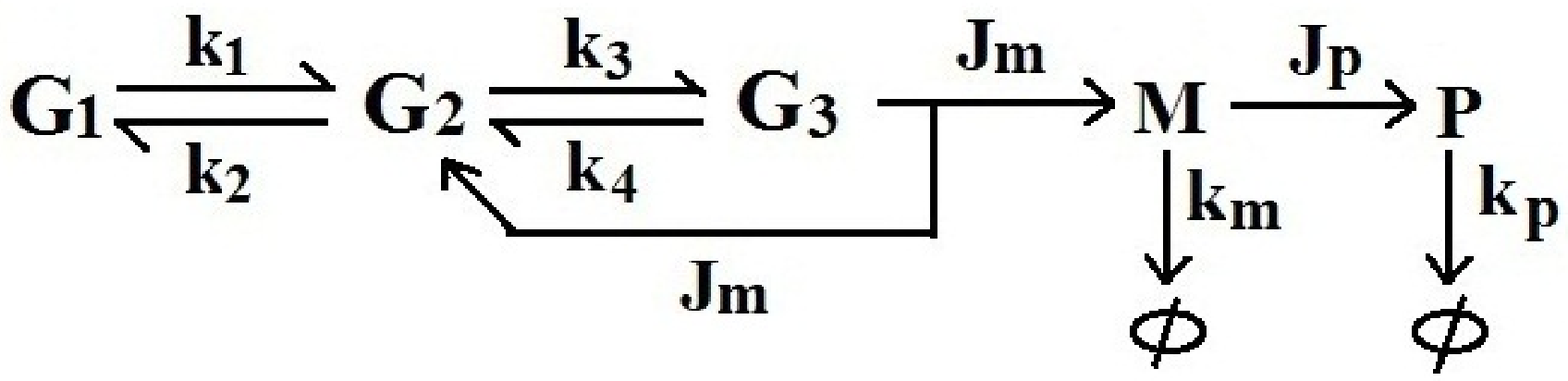} 
\par\end{centering}
Figure 1 Biochemical steps in gene expression with reinitiation and
the reaction rate constants. $k_{1}$($k_{2}$) is the activation
(inactivation) rate constant and $k_{3}$ is the rate constant for
the formation of initiation complex from the active state and $k_{4}$
is the rate constant for the dissociation of initiation complex. RNAP
molecules start transcription with rate constant $J_{m}$ from the
initiation complex. $J_{p}$ is the rate constant for protein synthesis
and $k_{m}$, $k_{p}$ are the rate constants for mRNA and protein
degradation respectively. 
\end{figure}

\subsection{Analysis}

Let us consider that there are $l$ copy number of a particular gene
exist in the cell. Let $p(l_{2},l_{3},m,n,t)$ be the probability
that at time $t$, there are $m$ number of mRNAs and $n$ number
of proteins molecules with $l_{2}$ number of genes in the active
state ($G_{2}$) and $l_{3}$ number of genes in the initiation complex
($G_{3}$). The number of genes in the inactive states are $l_{1}=(l-l_{2}-l_{3})$
as the gene copy number is conserved. The time evaluation of the probability
is given by the Master equation \cite{key-125}

\begin{equation}
\begin{array}{ccc}
\frac{\partial p(l_{2},l_{3},m,n,t)}{\partial t} & = & k_{1}[(l-l_{2}-l_{3}+1)p(l_{2}-1,l_{3},m,n,t)-(l-l_{2}-l_{3})p(l_{2},l_{3},m,n,t)]\\
 &  & +k_{2}[(l_{2}+1)p(l_{2}+1,l_{3},m,n,t)-l_{2}p(l_{2},l_{3},m,n,t)]\\
 &  & +k_{3}[(l_{2}+1)p(l_{2}+1,l_{3}-1,m,n,t)-l_{2}p(l_{2},l_{3},m,n,t)]\\
 &  & +k_{4}[(l_{3}+1)p(l_{2}-1,l_{3}+1,m,n,t)-l_{3}p(l_{2},l_{3},m,n,t)]\\
 &  & +J_{m}[(l_{3}+1)p(l_{2}-1,l_{3}+1,m-1,n,t)-l_{3}p(l_{2},l_{3},m,n,t)]\\
 &  & +k_{m}[(m+1)p(l_{2},l_{3},m+1,n,t)-mp(l_{2},l_{3},m,n,t)]\\
 &  & +J_{p}[mp(l_{2},l_{3},m,n-1,t)-mp(l_{2},l_{3},m,n,t)]\\
 &  & +k_{p}[(n+1)p(l_{2},l_{3},m,n+1,t)-np(l_{2},l_{3},m,n,t)]
\end{array}\label{eq:2.1}
\end{equation}

The standard generating function technique is now applied for the
expression of means and the variances of mRNA and protein in the steady
state \cite{key-125}.

The expressions for mean and FF of mRNAs ($<m>,\,FF_{m}$) and proteins
($<p>,\,FF_{p}$) in terms of the rate constants (figure. 1) at the
steady state are given by (for $l=1)$

\begin{equation}
<m>=\frac{k_{1}k_{3}}{a_{2}}\frac{J_{m}}{k_{m}},\qquad FF_{m}=1+\frac{J_{m}k_{3}(a_{2}-k_{1}a_{1})}{a_{2}(a_{1}k_{m}+a_{2})}\label{eq:2.4}
\end{equation}

\begin{equation}
<p>=\frac{<m>\,J_{p}}{k_{p}},\quad FF_{p}=1+\frac{J_{p}}{k_{p}+k_{m}}+\frac{J_{p}J_{m}a_{8}(k_{m}+a_{4})}{k_{m}\,(k_{p}+k_{m})\,a_{4}}+\frac{J_{p}J_{m}k_{3}(a_{6}-k_{1}k_{p}a_{8})}{k_{p}\,(k_{p}+k_{m})\,a_{5}}-\frac{J_{p}J_{m}k_{1}k_{3}}{k_{m}k_{p}\,a_{2}}\label{eq:2.5}
\end{equation}

where $a_{1}=k_{m}+J_{m}+k_{1}+k_{2}+k_{3}+k_{4},\,a_{2}=k_{1}J_{m}+k_{2}J_{m}+k_{2}k_{4}+k_{1}k_{3}+k_{1}k_{4},\,a_{3}=k_{m}+J_{m}+k_{4}$,
$a_{4}=k_{p}+J_{m}+k_{4}$, $a_{5}=a_{4}\{a_{2}+k_{p}(a_{1}+k_{p}-k_{m})\}$,
$a_{6}=(k_{p}a_{7}+k_{1})a_{4}+k_{p}a_{8}(J_{m}+k_{4})$, $a_{7}=a_{3}(k_{m}+k_{1})/(a_{1}k_{m}+a_{2})$
and $a_{8}=k_{3}(k_{m}+k_{1})/(a_{1}k_{m}+a_{2})$.

\begin{figure}[H]
\begin{centering}
\includegraphics[width=6cm,height=4cm]{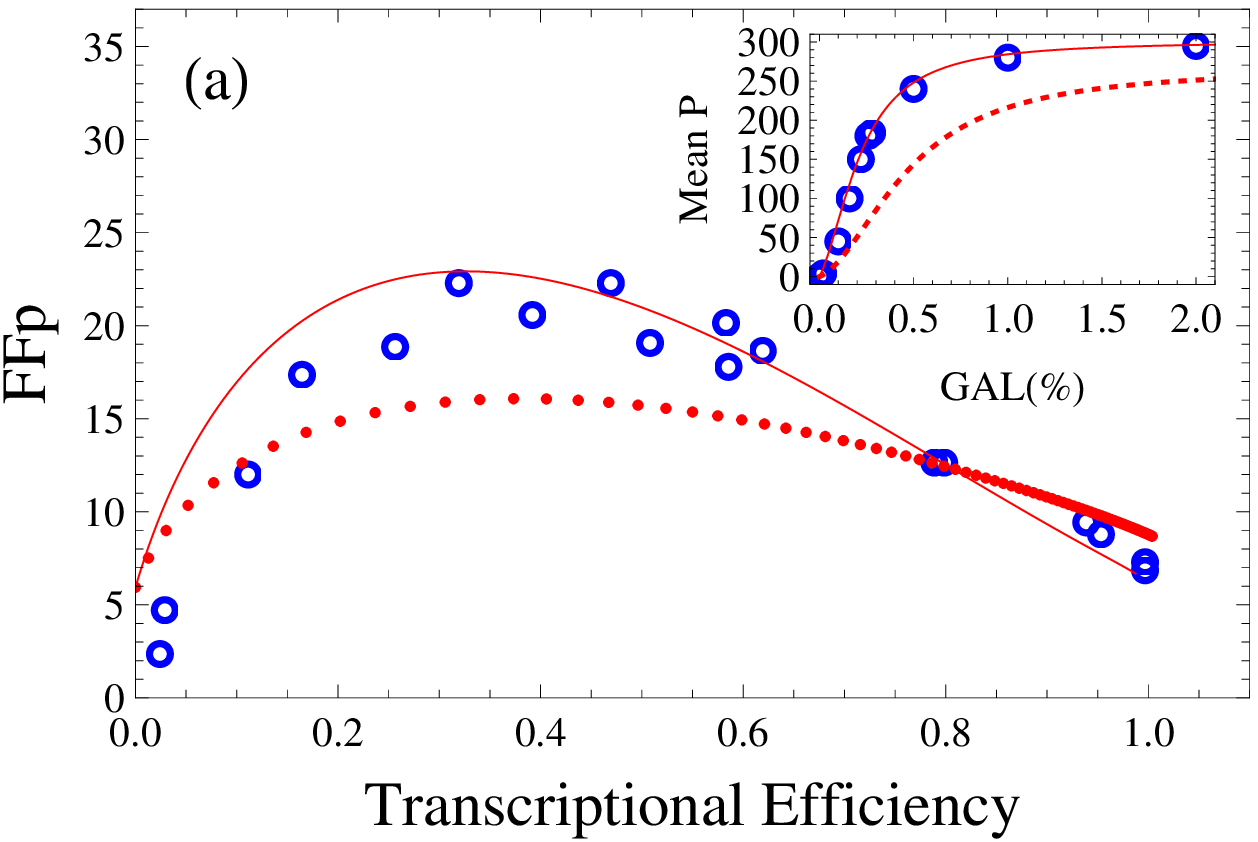} \includegraphics[width=5cm,height=4cm]{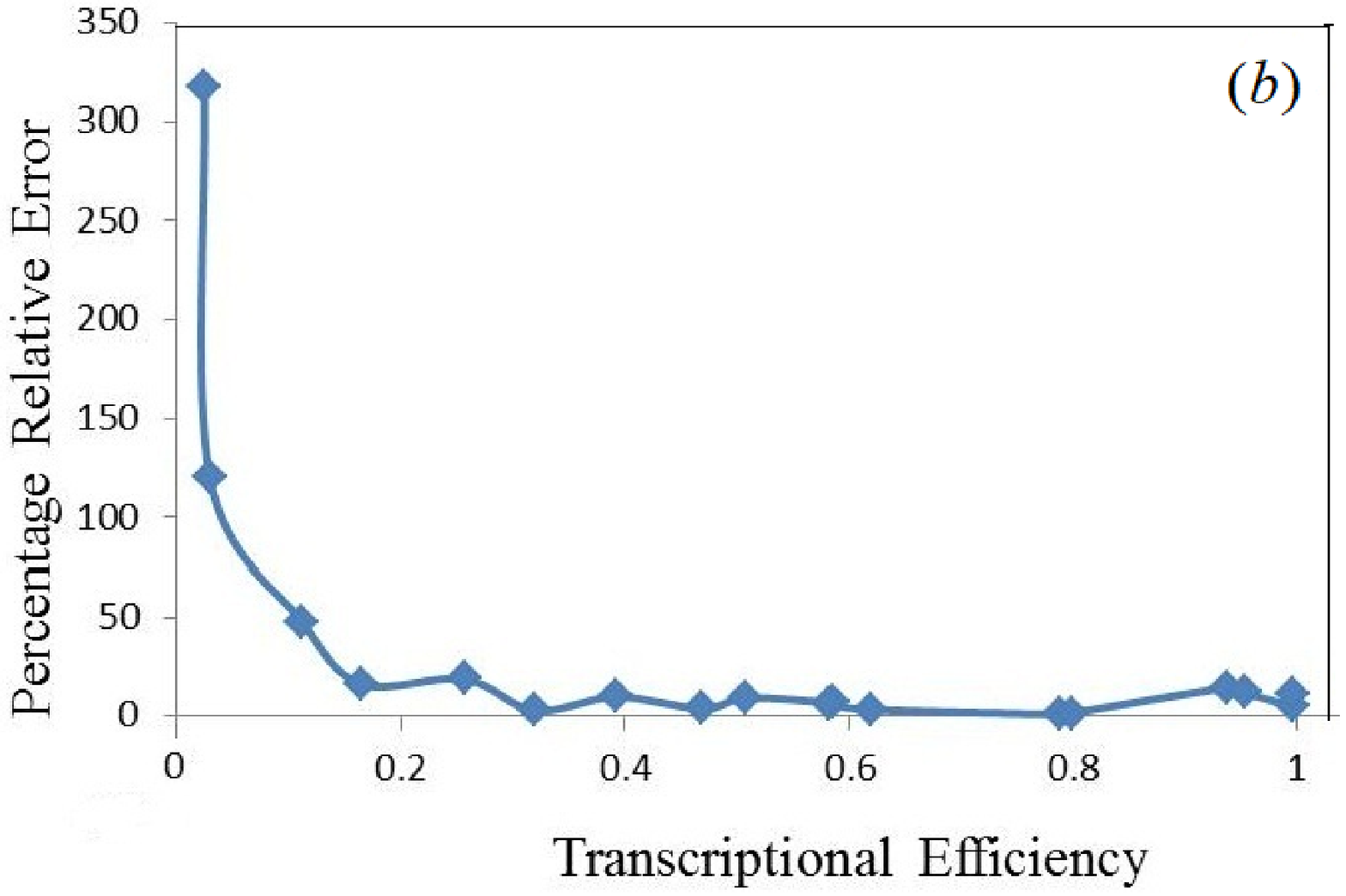} 
\par\end{centering}
Figure 2 (Colour online) \textbf{(a)} Plot of $FF_{p}$ with the transcriptional
efficiency. The rate constants are $k_{1}=0.02+0.2[GAL],$ $k_{2}=0.01+0.1[GAL]+0.077/[GAL],$
$k_{3}=50.0,$ $k_{4}=10.0,$ $J_{m}=1.0,$ $k_{m}=1.0,$ $J_{p}=5.0,$
$k_{p}=0.0125$ with reinitiation (red solid line) (figure 1) and
without reinitiation (red-dotted line) \cite{key-121}. The hollow
circles (blue) are the experimental data points with full ATC induction
\cite{key-111}. The inset figure shows the variation of mean protein
number with GAL concentration with reinitiation (red solid line) and
without reinitiation (red dotted line). \textbf{(b) }Plot of relative
percentage error of $FF_{p}$ with the transcriptional efficiency.
That shows the excellent agreement between the experimental data and
analytically calculated curve for intermediate and higher values of
transcriptional efficiency and a little disagreement at the lower
values of transcriptional efficiency. 
\end{figure}

Blake et al. measured the $FF_{p}$ as a function of transcriptional
efficiency by varying the inducer {[}GAL{]} (0 to 2\%) with full induction
of {[}ATC{]} (500ng/ml) and also by varying {[}ATC{]} with full induction
of {[}GAL{]} (2\%) \cite{key-111}. They found maximum $FF_{p}$ at
the intermediate values of transcriptional efficiency. We compared
our analytical result (eq.(\ref{eq:2.5})) with the experimental observations
of Blake et al. with full {[}ATC{]} induction. The plot of $FF_{p}$
with transcriptional efficiency along with the experimental data points
is shown in figure 2(a). We also plot the mean protein $<p>$ level
with GAL concentration (inset) with the same rate constants as given
in Blake et al. \cite{key-111}. We also plot the $FF_{p}$ with transcriptional
efficiency and $<p>$ with GAL concentration (inset) without the reinitiation
of transcription by RNAP (red dotted curves in figure 2(a)). We plot
the percentage relative error between the experimental data points
and analytically calculated values with transcriptional efficiency
in figure 2(b). That shows the excellent agreement between the experimental
data in Blake et al. \cite{key-111} and our analytically calculated
curve for intermediate and higher values of transcriptional efficiency
and a little disagreement at the lower values of transcriptional efficiency.
The initial disagreement may be due to the presence of repressors
in the experimental system. We also observe the large deviation of
$FF_{p}$ and $<p>$ from the experimental data points without the
reinitiation of transcription by RNAP.

The transcriptional reinitiation by RNAP affects the FF at mRNA level
first. Post transcription processes then further amplify that effect.
So, we study the $FF_{m}$ to explore the role of reinitiation of
transcription. In the expression of $FF_{m}$ in equation (\ref{eq:2.4}),
the first term arises due to the random birth and death of mRNA molecules
and the second term arises due to the random transitions between different
gene states, $G_{1}$, $G_{2}$ and $G_{3}$ (figure 1). Eq. (\ref{eq:2.4})
shows that the reinitiation can drive the $FF_{m}$ to the sub-Poissonian
level depending on the condition, given by

\begin{equation}
a_{2}<k_{1}a_{1}\;or\;(J_{m}+k_{4})<\frac{k_{1}}{k_{2}}(k_{1}+k_{2}+k_{m})\label{eq:2.6}
\end{equation}

And there is a critical value $J_{m}^{c}$ for the given value of
$k_{1}$, $k_{2}$ and $k_{4}$ and that is given by

\begin{equation}
J_{m}^{c}=\frac{k_{1}}{k_{2}}(k_{1}+k_{2}+1)-k_{4}\label{eq:2.7}
\end{equation}

The transcriptional reinitiation process in figure 1 is mainly controlled
by the rate constants $k_{3}$ and $J_{m}.$ We observe the variation
of $FF_{m}$ with $k_{3}$ for the rate constants $k_{1}=6.02$ and
$k_{2}=3.012$ (i.e., high $GAL$ concentration ($30\%$)) with $J_{m}$
as parameter (figure 3(a)). For the rate constant considered in figure
3, the critical value of $J_{m}$ is $J_{m}^{c}=10.048$ for which
the $FF_{m}$ is Poissonian (green, solid line in figure 3(a)). For
$J_{m}$ greater than $J_{m}^{c}$, the $FF_{m}$ always lies in the
super-Poissonian phase (blue dashed line for $J_{m}=13.048$ in figure
3(a)) whereas for $J_{m}$ less than $J_{m}^{c}$ the $FF_{m}$ always
lies in the sub-Poissonian phase (red dotted line for $J_{m}=1.048$
in figure 3(a)). The green solid line ($FF_{m}$=1) in figure 3(a)
clearly separates the sub-Poissonian and super-Poissonian phases.
The degree of deviation of $FF_{m}$ towards sub-Poissonian phase
increases with the decreases in $k_{2}$ and $k_{4}$. The variation
of $FF_{p}$ is plotted against $k_{3}$ with above mentioned three
different values of $J_{m}$. Translation in gene expression produces
burst of proteins from each mRNAs. That bursting process adds some
extra amount of fluctuations in protein levels (eq.(\ref{eq:2.5})).
The straight solid line (for $J_{m}=10.048$) and curved dot-dashed
line (for $J_{m}=8.048$) in figure 3(a) becomes curved and straight
respectively in figure 3(b) due to translation in protein synthesis.
In the previous models and calculations, the $FF_{m}$ is always found
to be greater than unity \cite{key-115,key-121,key-124A,key-124B}.
But, with the inclusion of reinitiation processes in gene transcription,
we observe that two distinct phases viz. super-Poissonian and sub-Poissonian
Fano factors separated by Poissonian one are possible. Depending on
the values of different rate constants, the $FF_{m}$ can be in any
one of the phases.

\begin{figure}[H]
\begin{centering}
\includegraphics[width=5cm,height=3cm]{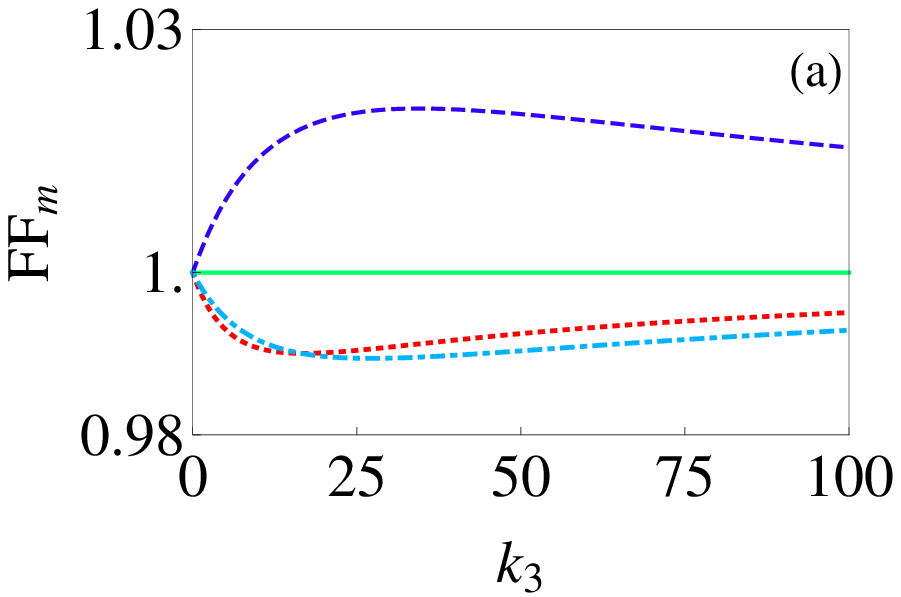} \includegraphics[width=5cm,height=3cm]{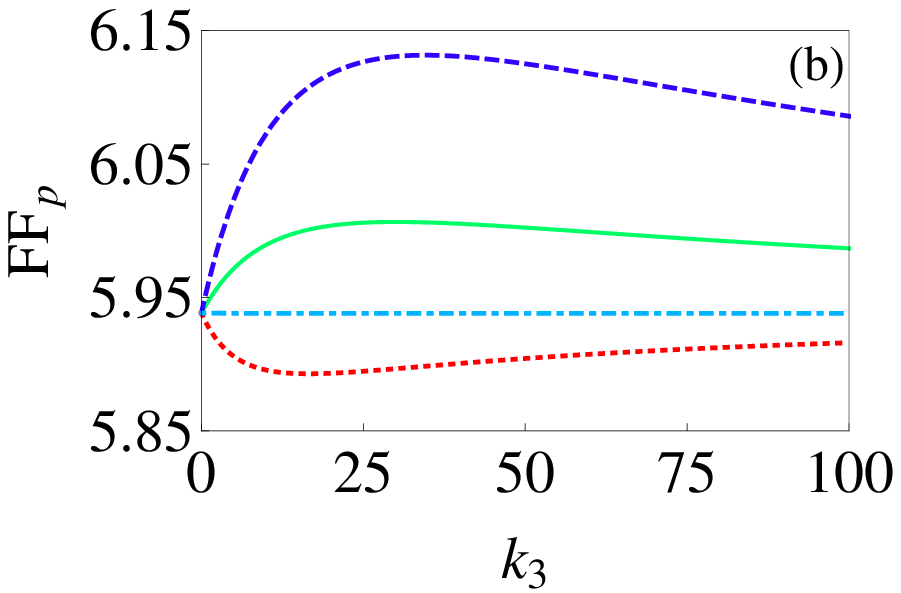} 
\par\end{centering}
Figure 3 (Color online) Plot of Fano factors ((a) $FF_{m}$ and (b)
$FF_{p}$) with the rate constant $k_{3}$ for $30\%$ $GAL$ concentration
with $J_{m}$ as parameter. Other rate constants are same as in figure
1(b). Four different values of $J_{m}$ are: $1.048$ (Red dotted),
$8.048$ (Cyan dot-dashed), $10.048$ (Green solid) and $13.048$
(Blue dashed). 
\end{figure}

\begin{figure}[H]
\begin{centering}
\includegraphics[width=5cm,height=3cm]{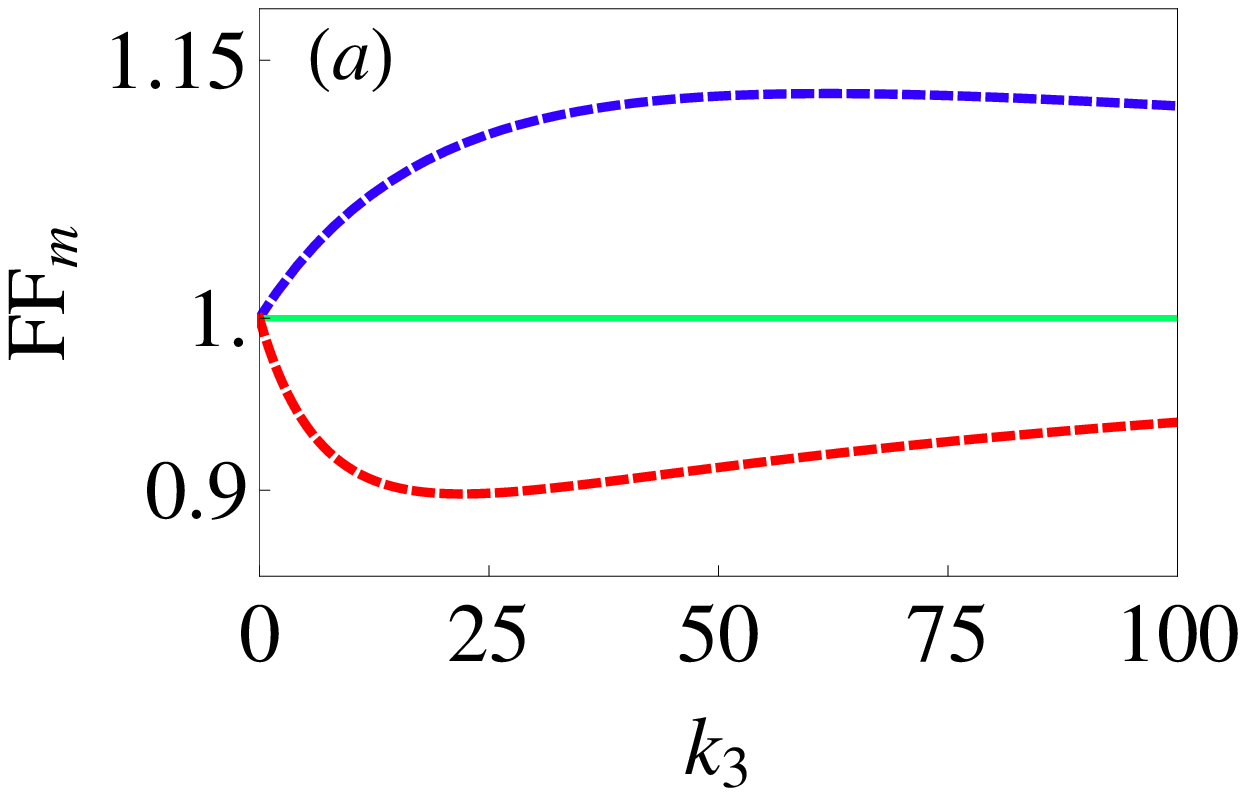}\includegraphics[width=5cm,height=3cm]{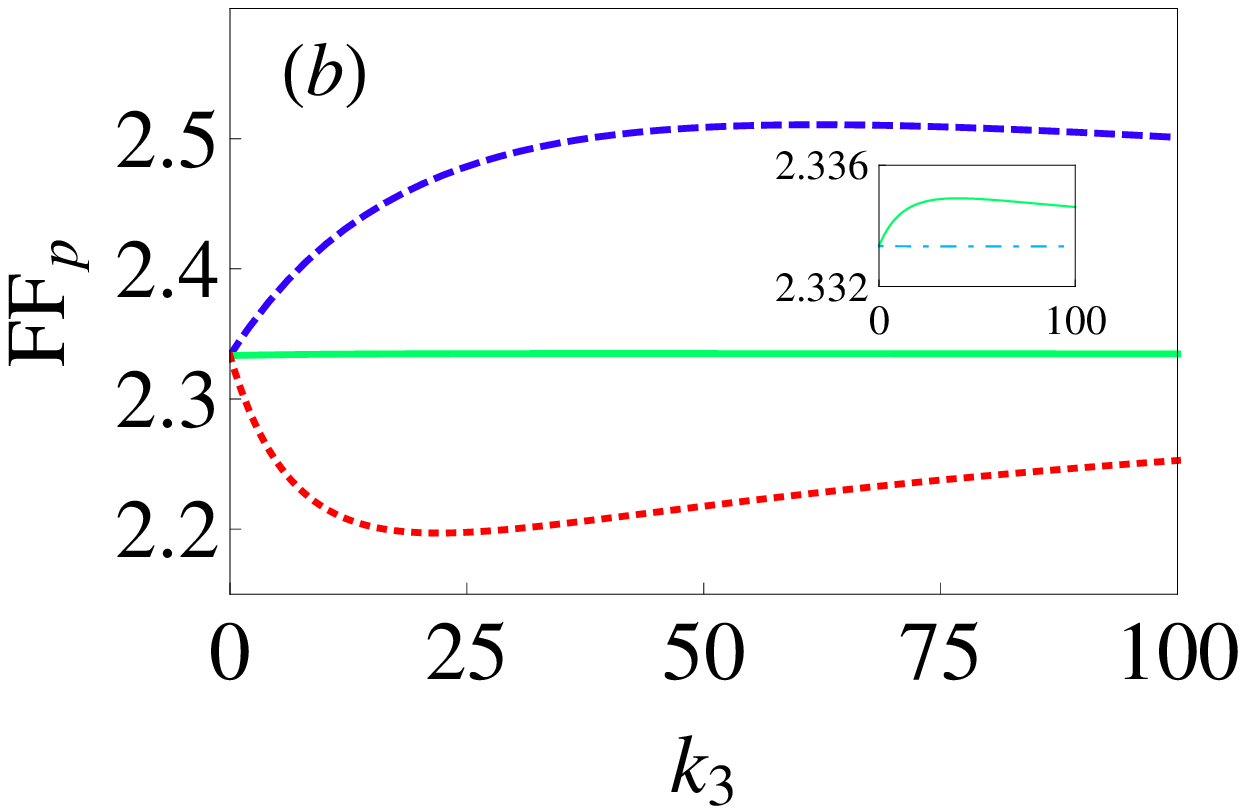} 
\par\end{centering}
\begin{centering}
\includegraphics[width=5cm,height=3cm]{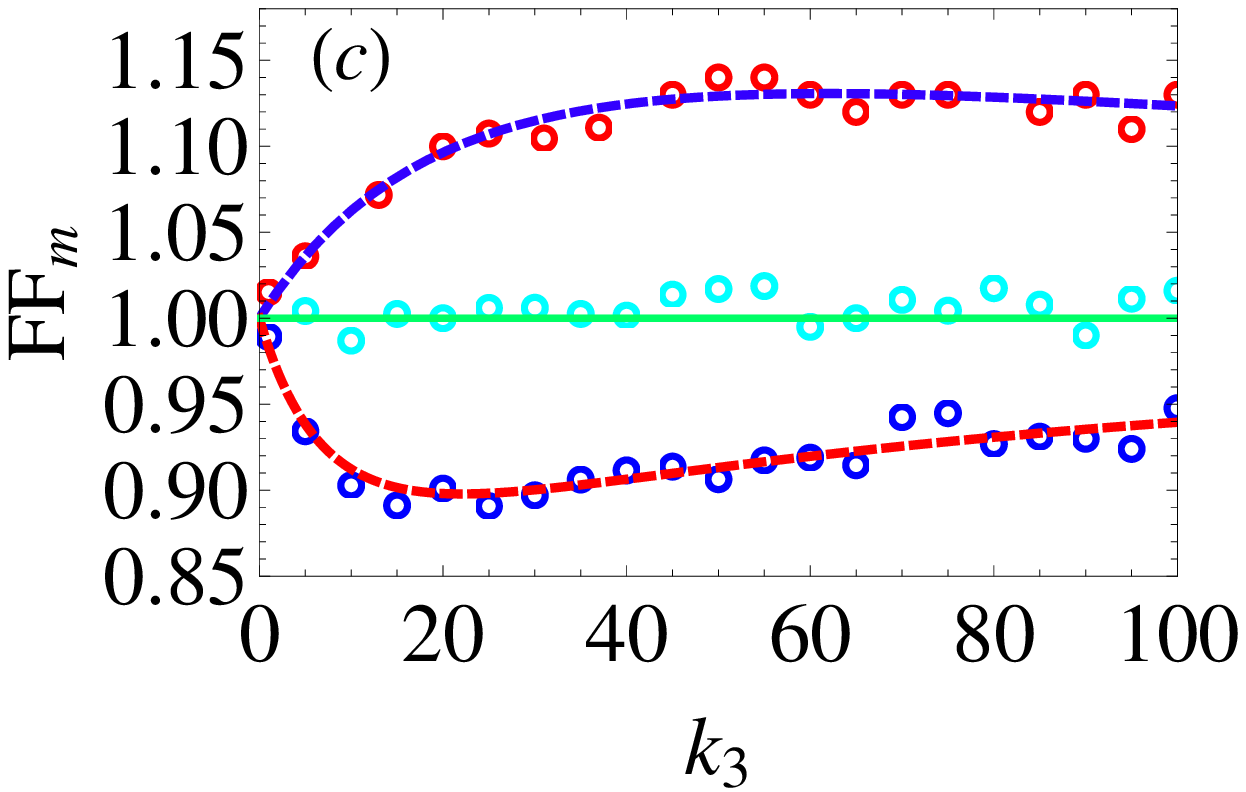} \includegraphics[width=5cm,height=3cm]{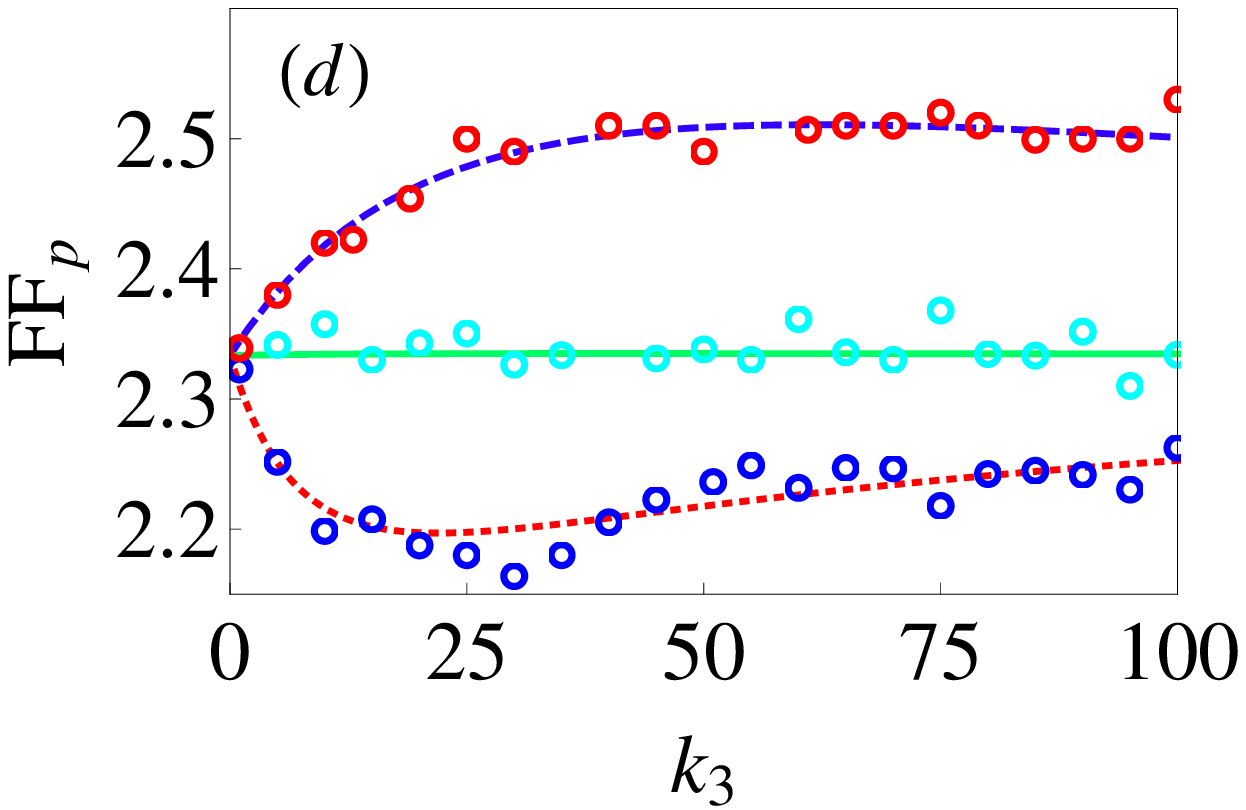} 
\par\end{centering}
Figure 4 (Color online) Plot of Fano factors ((a) $FF_{m}$ and (b)
$FF_{p}$) with the rate constant $k_{3}$ and $J_{m}$ as parameter.
Other rate constants are $k_{1}=10.0,\;k_{2}=10.0,\;k_{4}=1.0,\,k_{m}=0.1,$
$J_{p}=0.2$ and $k_{p}=0.05.$ The rate constants are taken from
Kaern et al. \cite{key-115} except $k_{4}$ which is chosen arbitrarily.
The value of $J_{m}$ are : $30.0$ (Blue dashed), $19.1$ (Green
solid) and $10.0$ (Red dotted). $J_{m}=19.0$ for the (cyan dash-dot)
straight curve in the inset (b). The hollow circles in (c) and (d)
are generated from stochastic simulation using Gillespie algorithm
corresponding to the curves in (a) and (b). 
\end{figure}

The rate constants for gene expression for different gene are not
unique. The different literature on gene expression pointed out different
values of rate constants for different steps. Figures 4(a) and 4(b)
show the more intense variation of $FF_{m}$ with $k_{3}$ for the
rate constants given in Kaern et al. \cite{key-115}. We also calculate
the $FF_{m}$ and $FF_{p}$ from stochastic simulation using Gillespie
algorithm with the rate constants corresponding to the curves in figures
4(a) and 4(b) \cite{key-125A}. Figures 4(c) and 4(d) show that our
analytical results agree well with the simulation results.

To observe the more clear variation of $FF_{m}$ with other rate constants
over wide range, we use surface plots with different rate constants
(figures 5-7). Figures 5-7 show that higher values of $k_{1}$ and
lower values of $k_{2}$, $k_{4}$ and $J_{m}$ favour sub-Poissonian
$FF_{m}$. At a very high value of $k_{1}$, the $FF_{m}$ becomes
independent of $k_{2}$ (figure 6(a)). We see from eq.(\ref{eq:2.7})
that the critical value of $J_{m}$ increases with the increase of
rate constant $k_{1}$ and decrease with rate constants $k_{2}$ and
$k_{4}$. That is reflected in figures 5(b) and 7(b). We observe from
eq.(\ref{eq:2.4}) that $k_{1}$, $k_{3}$ and $J_{m}$ play crucial
role for the determination of mean mRNA level. Figure 8 shows that
the mean mRNA level does not increases much with the increase of $k_{1}$
and $k_{3}$ but increases more with $J_{m}$. Therefore, the variation
of $k_{1}$ or $k_{3}$ is sufficient to change the mean mRNA level
very little but the variation of $J_{m}$ is necessary for more change
in mRNA level. This behaviour is similar like 'coarse' and 'fine'
control knob of electronic devices. Now, if $k_{1}$ and $J_{m}$
are increased (keeping $J_{m}<J_{m}^{c}$) simultaneously to increase
the mean mRNA number, the $FF_{m}$ moves to the sub-Poissonian regime
(figures 5(b) and 8(b)).

In the experiment of Blake et al., the rate constants $k_{1}$ and
$k_{2}$ are function of GAL concentrations. The rate constant $k_{1}$
increases monotonically with the increase of GAL concentration but
the rate constant $k_{2}$ initially decreases (up to 3\%) and then
increases with the increase of GAL concentration (beyond 3\%). Figure
9(a) shows that average mRNA level does not increase with the increase
of GAL concentration beyond 2\%. In figure 9(b), we see that the $FF_{m}$
can be sub-Poissonian with different values of GAL concentration with
the given reinitiation rate constant $J_{m}$. Higher GAL concentration
is required for sub-Poissonian $FF_{m}$ for higher $J_{m}$. That
is, the $FF_{m}$ can be decreased without changing the mean mRNA
level by varying GAL concentration only (keeping other rate constants
fixed). Sanchez et al. observed that the FF can be decreased substantially
with fixed mean by varying {[}GAL{]} only beyond 2\% in GAL1{*} promoter
\cite{key-124A}.

\begin{figure}[H]
\begin{centering}
\includegraphics[width=5cm,height=3cm]{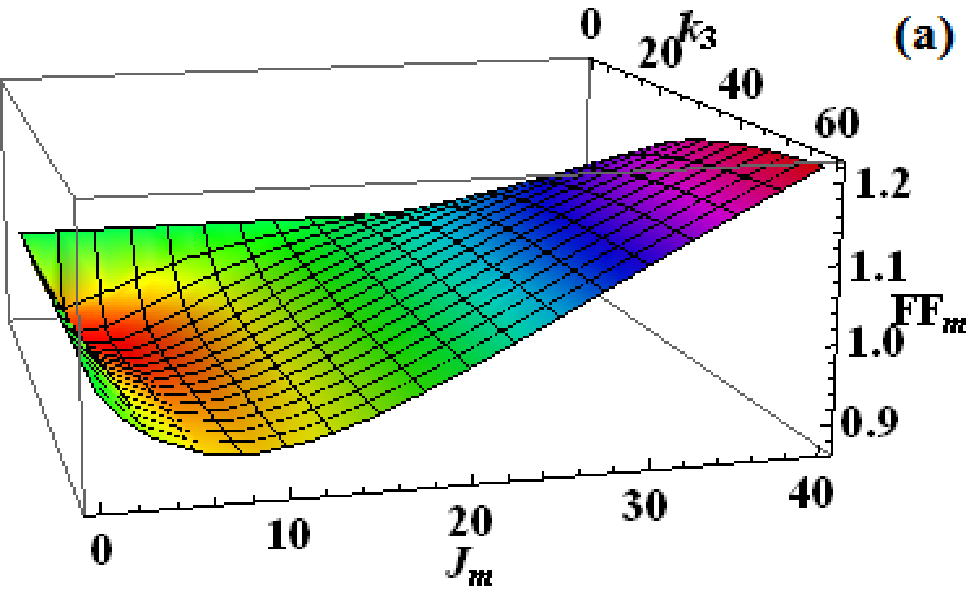} \includegraphics[width=5cm,height=3cm]{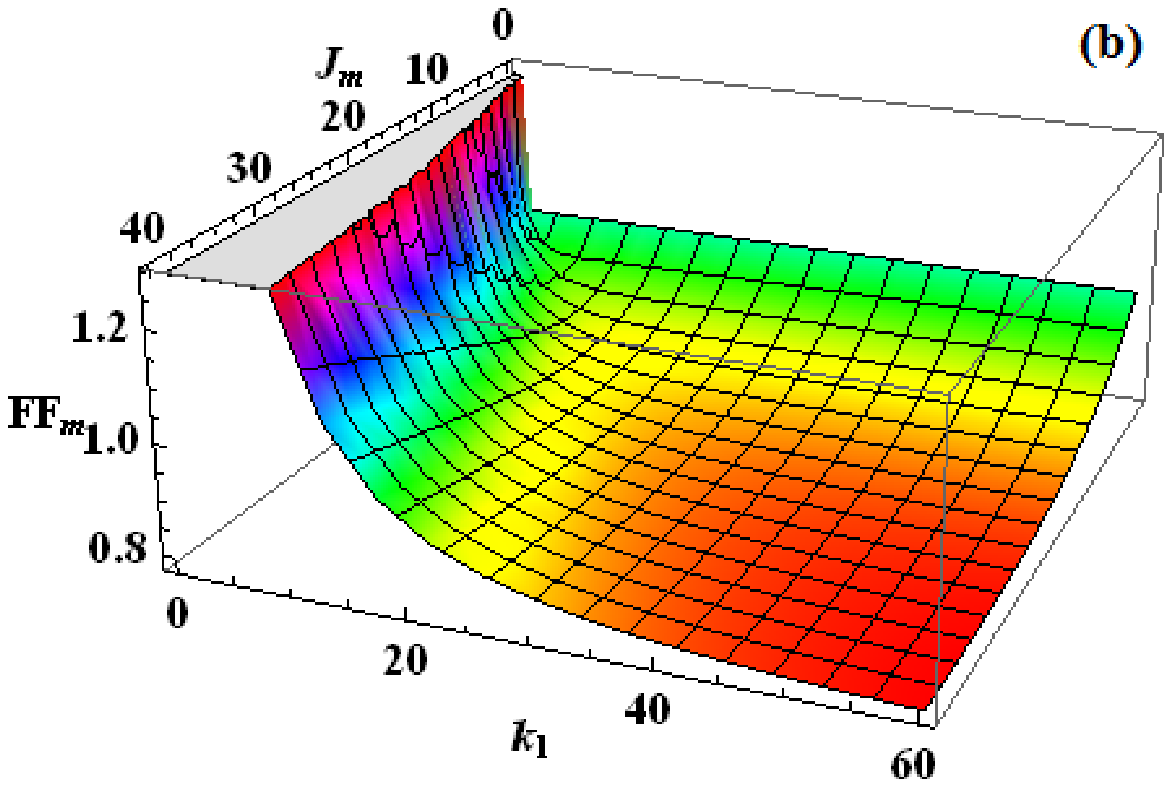} 
\par\end{centering}
Figure 5 (Color online) The variation of $FF_{m}$ with (a) $J_{m}$
and $k_{3}$ for the rate constants $k_{1}=10.0,\;k_{2}=10.0,\;k_{4}=1.0$.
and with (b) $J_{m}$ and $k_{1}$ for $k_{2}=10.0,\:k_{3}=50.0,\;k_{4}=1.0$. 
\end{figure}

\begin{figure}[H]
\begin{centering}
\includegraphics[width=5cm,height=3cm]{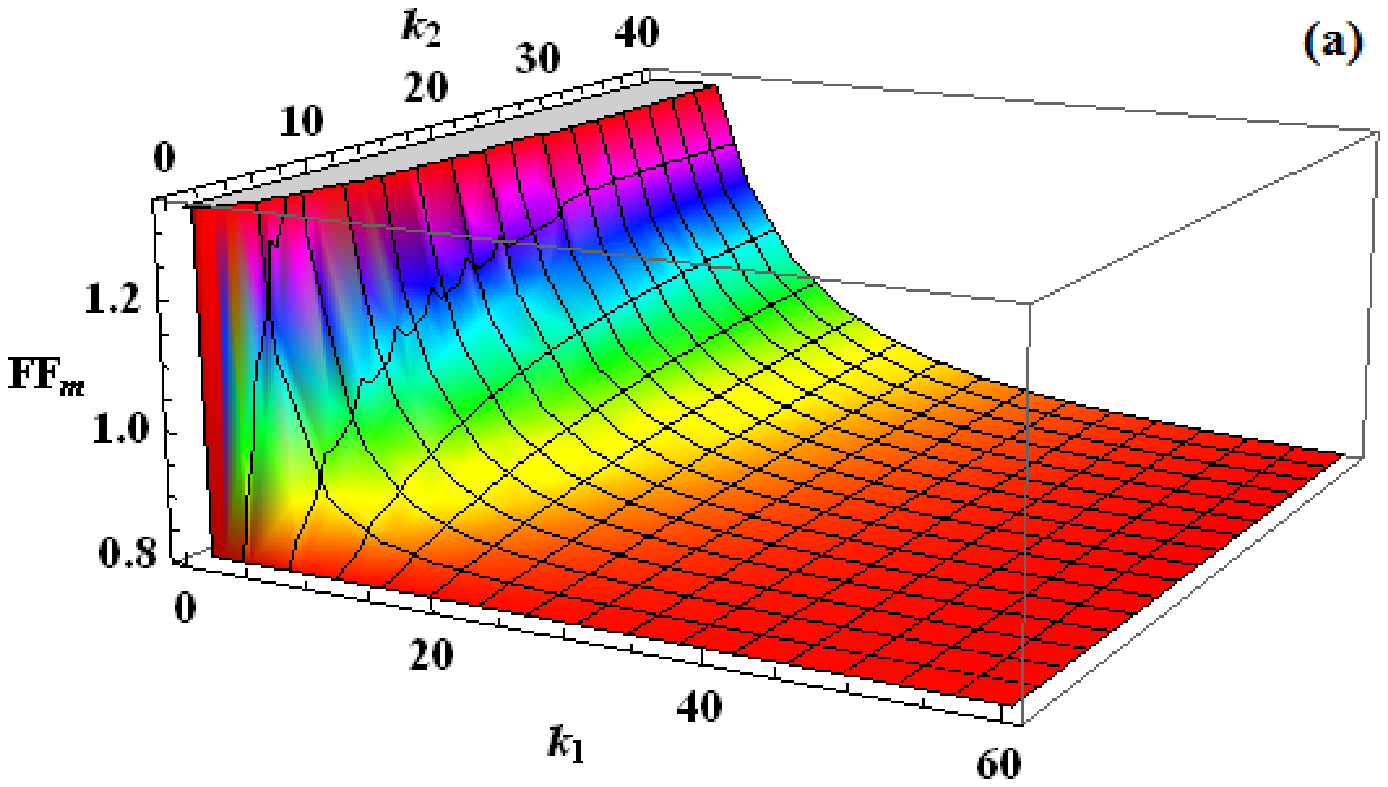}\includegraphics[width=5cm,height=3cm]{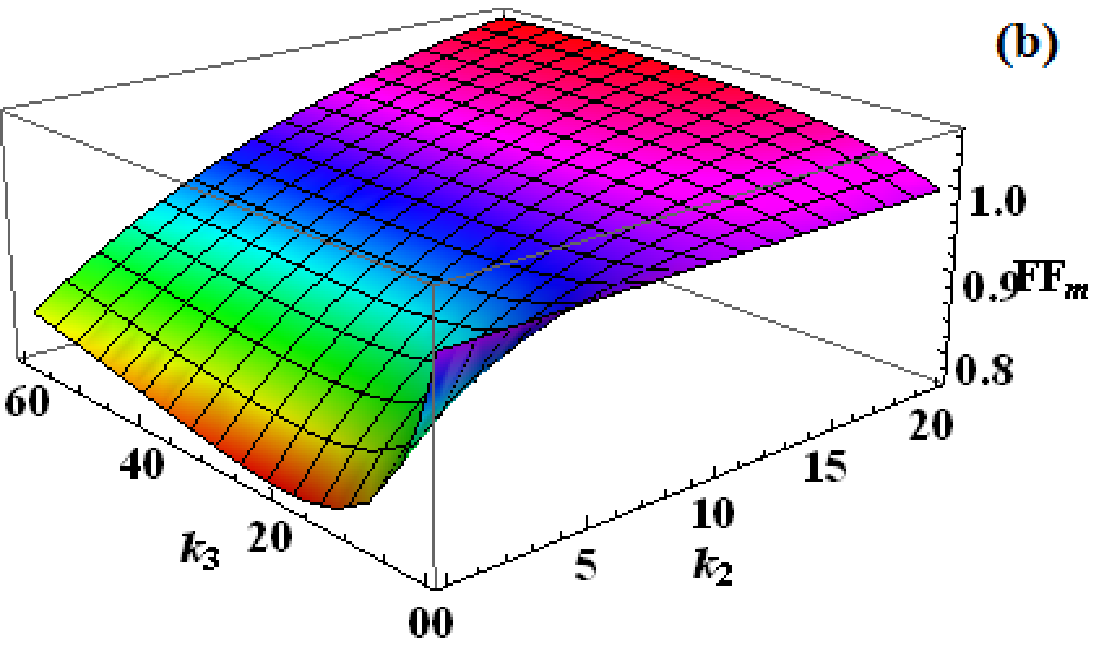} 
\par\end{centering}
Figure 6 (Color online) The variation of $FF_{m}$ with (a) $k_{1}$
and $k_{2}$ for the rate constants $k_{3}=50,$ $k_{4}=1$, $J_{m}=19$
and with (b) $k_{2}$ and $k_{3}$ for the rate constants $k_{1}=10,$
$k_{4}=1$, $J_{m}=19$. 
\end{figure}

\begin{figure}[H]
\begin{centering}
\includegraphics[width=5cm,height=3cm]{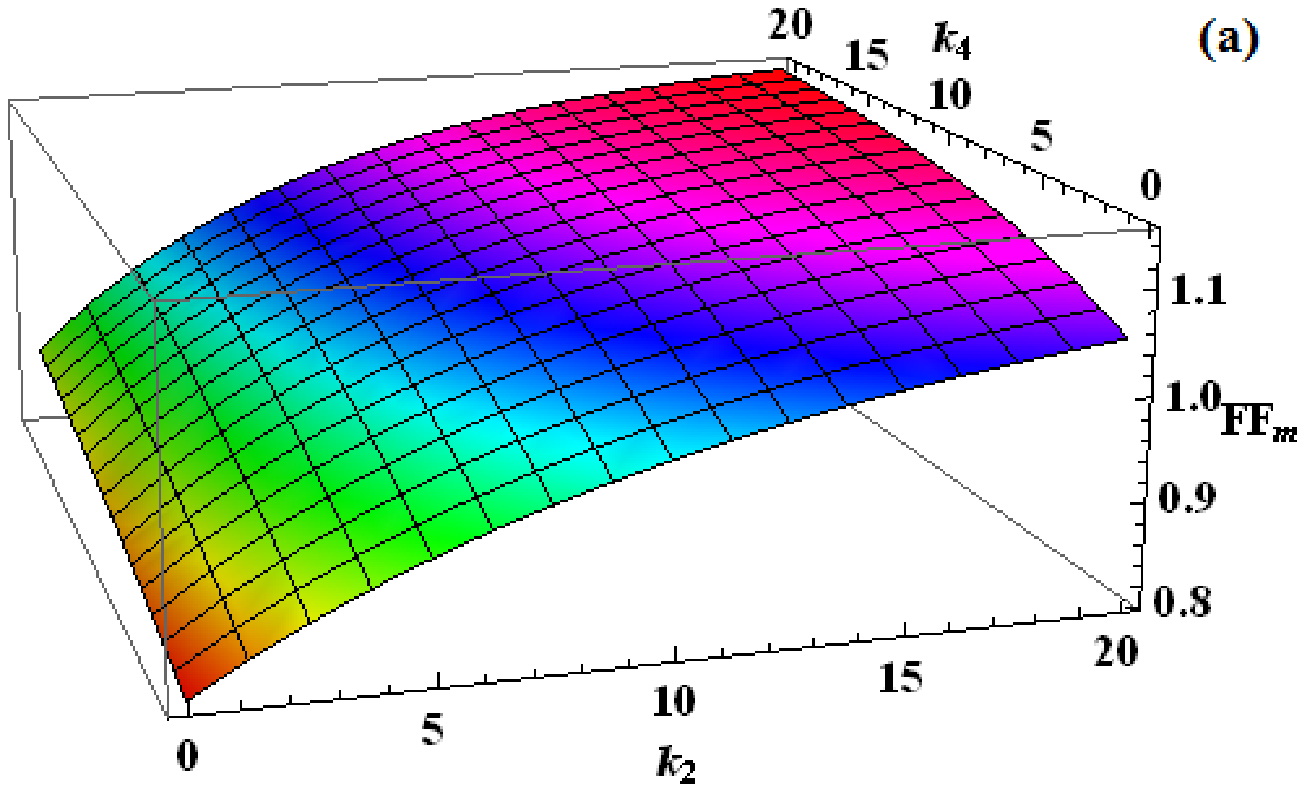}\includegraphics[width=5cm,height=3cm]{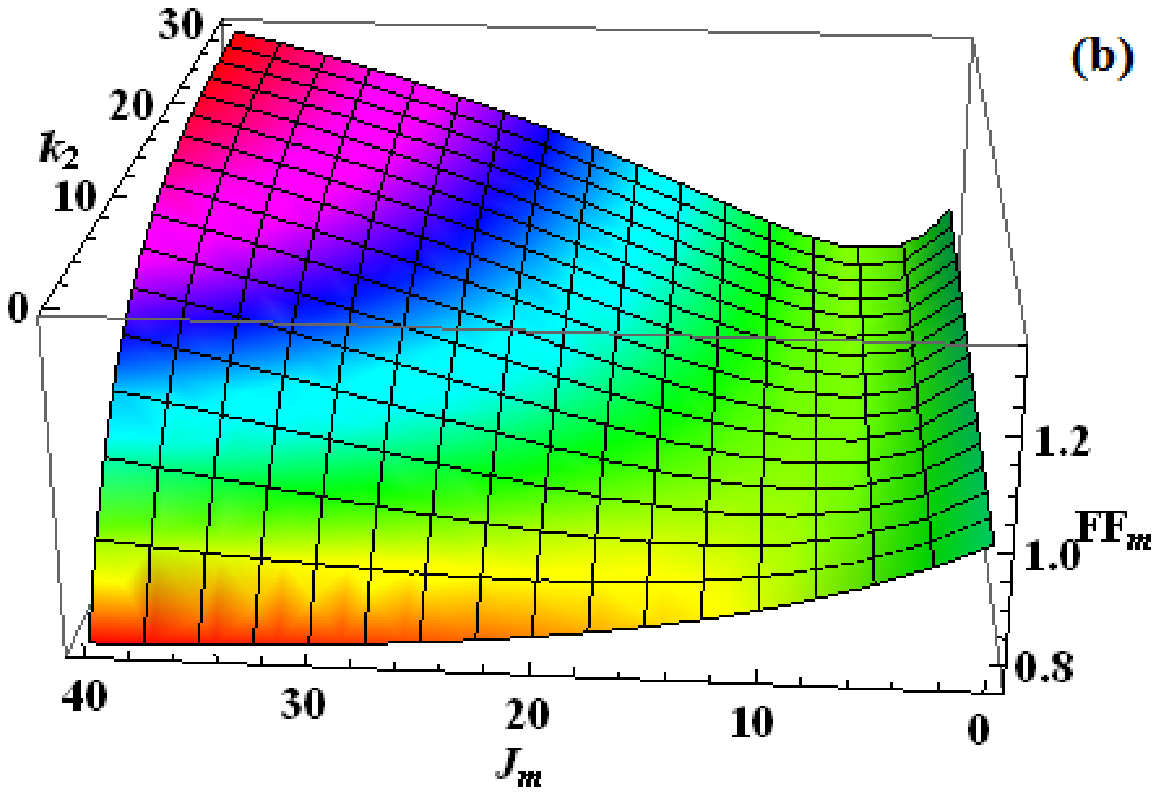} 
\par\end{centering}
Figure 7 (Color online) The variation of $FF_{m}$ with (a) $k_{2}$
and $k_{4}$ with other rate constants as $k_{3}=50,$ $k_{1}=10$,
$J_{m}=19$ and with (b) $J_{m}$ and $k_{2}$ with other rate constants
as $k_{1}=10,$ $k_{3}=50$, $k_{4}=1$. 
\end{figure}

\begin{figure}[H]
\begin{centering}
\includegraphics[width=5cm,height=3cm]{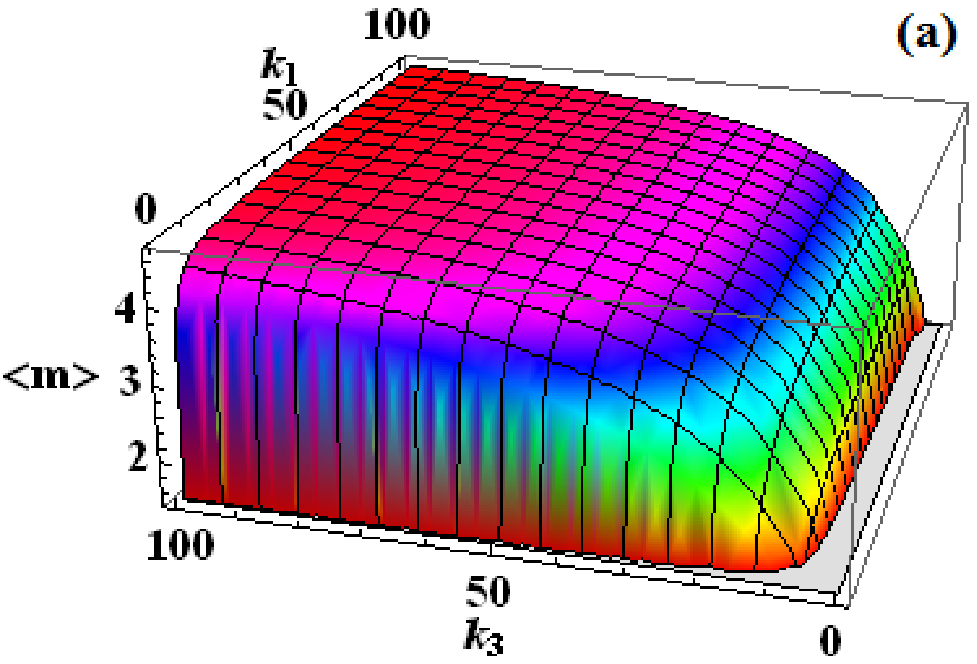} \includegraphics[width=5cm,height=3cm]{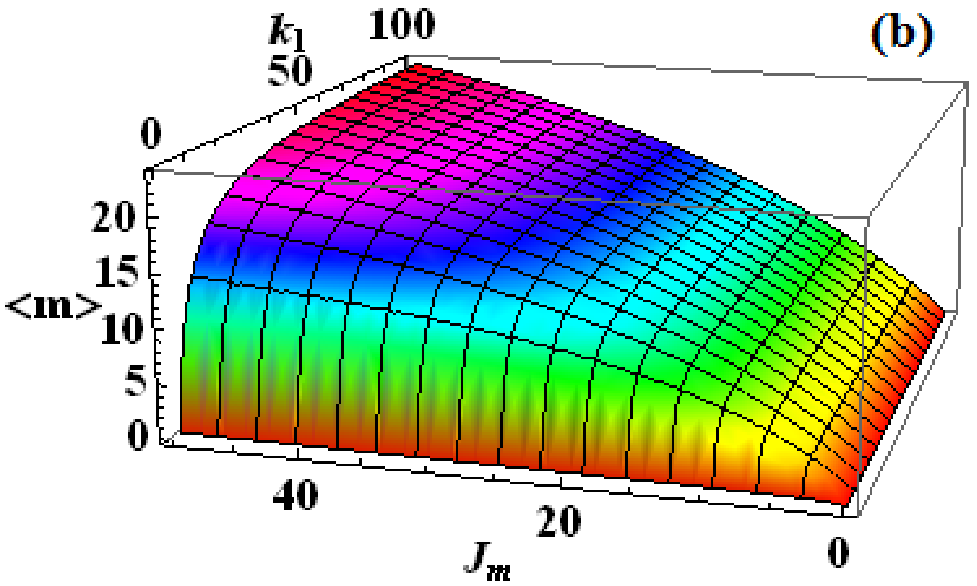}\includegraphics[width=5cm,height=3cm]{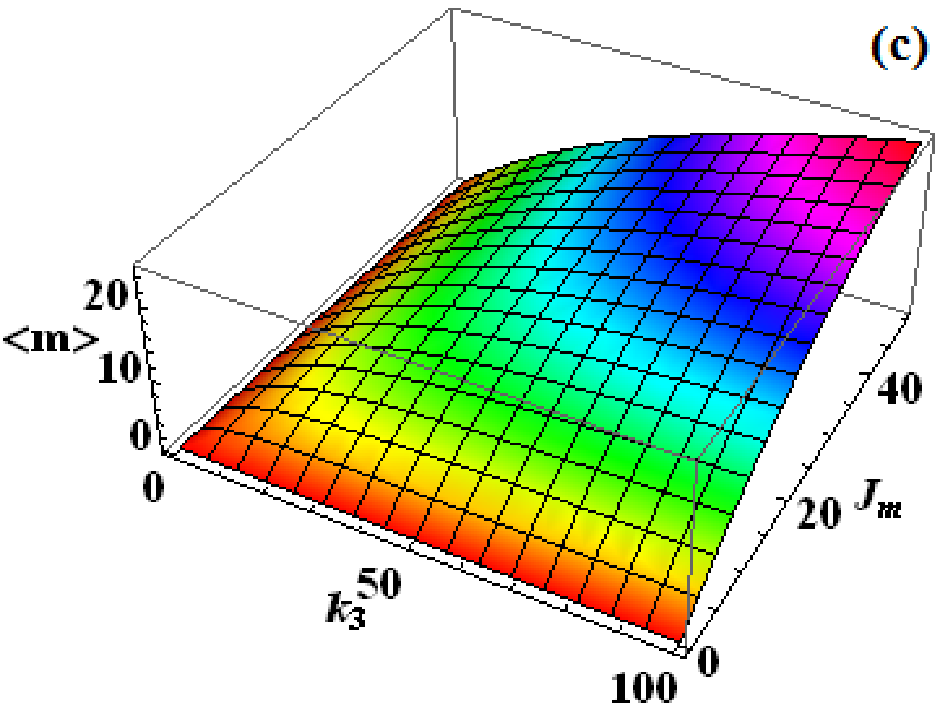} 
\par\end{centering}
Figure 8 (Color online) Mean mRNA number varies with (a) $k_{1}$
and $k_{3}$ (other rate constants are $J_{m}=5,$$k_{2}=10$, $k_{4}=1$)
(b) $k_{1}$ and $J_{m}$ (other rate constants are $k_{2}=10,$ $k_{3}=50$,
$k_{4}=1$) and (c) $k_{3}$ and $J_{m}$ (other rate constants are
$k_{1}=10,$ $k_{2}=10$, $k_{4}=1$). 
\end{figure}

\begin{figure}[H]
\begin{centering}
\includegraphics[width=5cm,height=3cm]{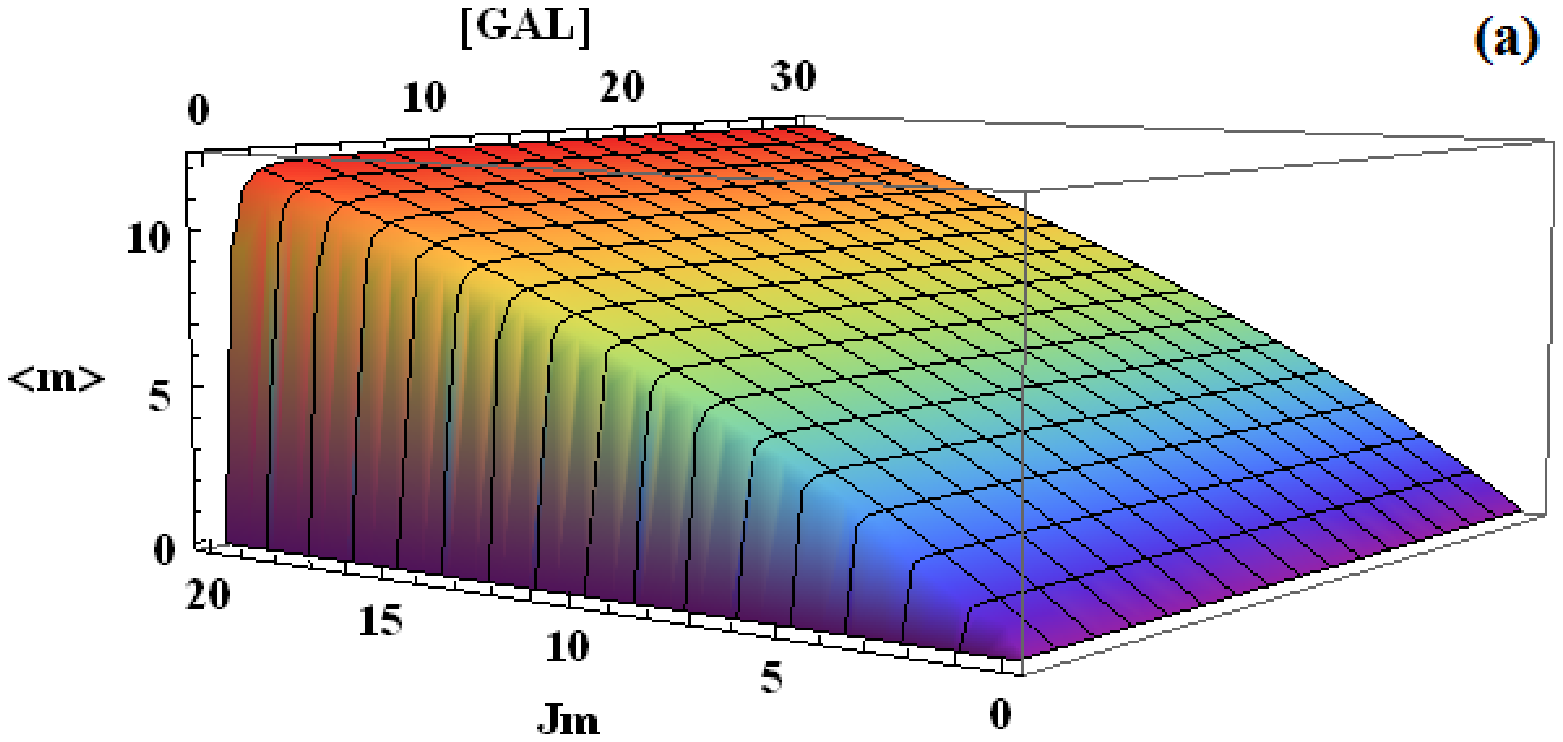}\includegraphics[width=5cm,height=3cm]{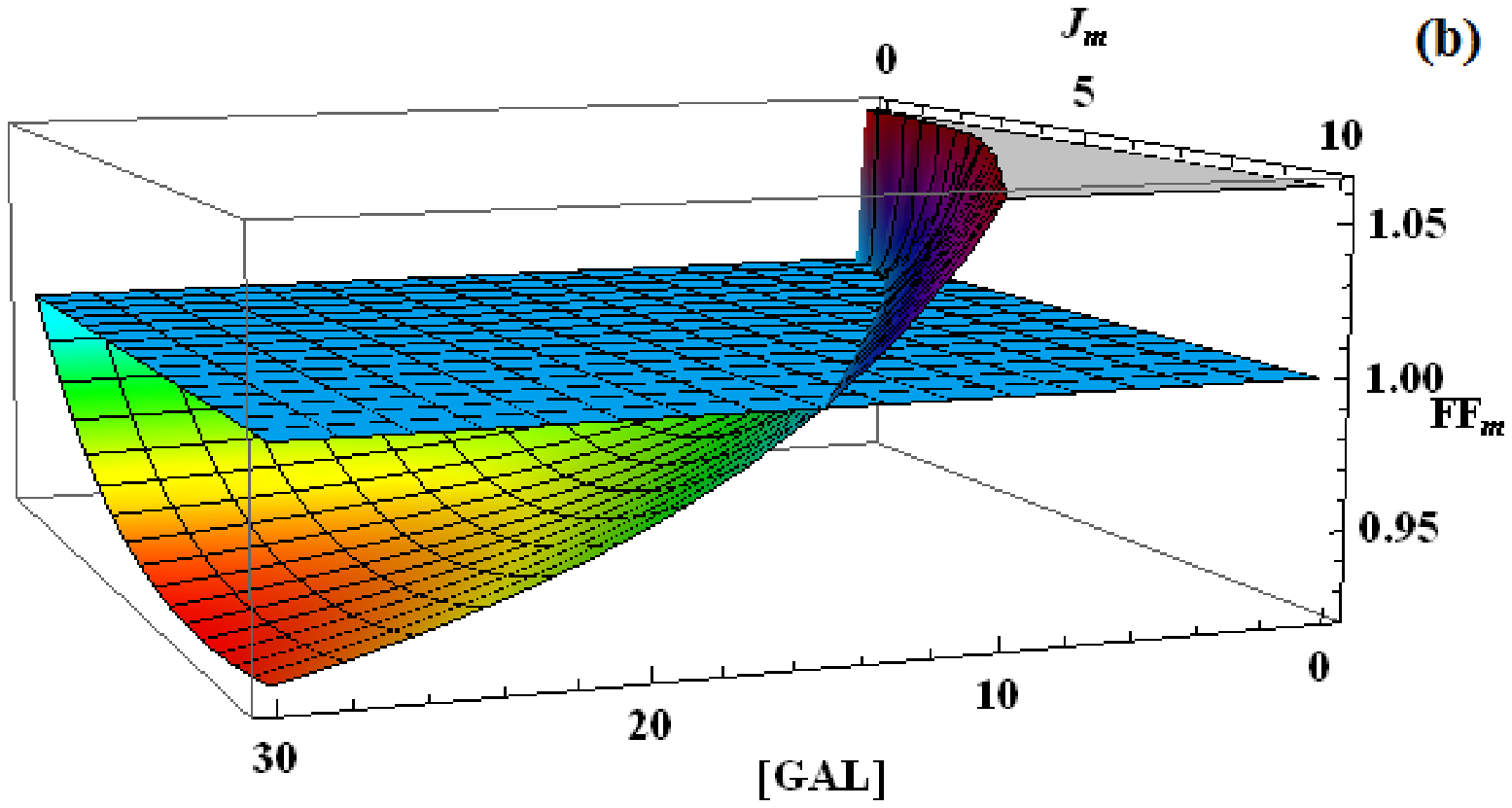} 
\par\end{centering}
Figure 9 (Color online) (a) Variation of mean mRNA ($<m>$) and (b)
$FF_{m}$ with {[}GAL{]} and $J_{m}$. The other rate constants are
($k_{3}=50.0,$ $k_{4}=1.0$). 
\end{figure}

\section{Steady state probability density function for mRNA}

For a single copy number of the gene ($l=1$), the equation (\ref{eq:2.1})
can be decomposed into three different equations corresponding to
the three different gene states from which the

\begin{equation}
\frac{\partial p_{1}(m,t)}{\partial t}=k_{2}p_{2}(m,t)-k_{1}p_{1}(m,t)+k_{m}[(m+1)p_{1}(m+1,t)-mp_{1}(m,t)]\label{eq:2.9}
\end{equation}

\begin{equation}
\frac{\partial p_{2}(m,t)}{\partial t}=k_{1}p_{1}(m,t)+k_{4}p_{3}(m,t)-(k_{2}+k_{3})p_{2}(m,t)+J_{m}p_{3}(m-1,t)+k_{m}[(m+1)p_{2}(m+1,t)-mp_{2}(m,t)]\label{eq:2.10}
\end{equation}
\begin{equation}
\frac{\partial p_{3}(m,t)}{\partial t}=k_{3}p_{2}(m,t)-k_{4}p_{3}(m,t)-J_{m}p_{3}(m,t)+k_{m}[(m+1)p_{3}(m+1,t)-mp_{3}(m,t)]\label{eq:2.11}
\end{equation}
where $p_{i}(m,t)$ ($i=1,2,3$) be the probability that at time $t,$
gene is in the $G_{i}$($i=1,2,3$) state and the number of $mRNA$
molecule is $m$. 
\begin{flushleft}
Now using the standard generating function technique \cite{key-118,key-125},
we have the total probability density function from the Master equations
(\ref{eq:2.9}) and (\ref{eq:2.10}) 
\par\end{flushleft}

\begin{equation}
p(m)=\frac{(-h_{4})^{m}\,\Gamma(h_{1}+m)\,\Gamma(h_{2}-h_{3})\,\Gamma(h_{2}+h_{3})}{\Gamma(m+1)\,\Gamma(h_{1})\,\Gamma(h_{2}-h_{3}+m)\,\Gamma(h_{2}+h_{3}+m)}\,_{p}F_{q}[\{h_{1}+m\},\{h_{2}-h_{3}+m,\,h_{2}+h_{3}+m\},\,h_{4}]\label{eq:2.8}
\end{equation}

where $h_{1}=\frac{s_{4}}{s_{3}}=k_{1}$ $h_{2}=-{\displaystyle \frac{1}{2}+{\displaystyle \frac{s_{1}}{2}}}$,
$h_{3}={\displaystyle \frac{1}{2}}{\displaystyle \sqrt{1-2s_{1}+s_{1}^{2}-4s_{2}}}$,
$h_{4}=s_{3}$ and $s_{1}=1+J_{m}+k_{1}+k_{2}+k_{3}+k_{4}$, $s_{2}=k_{1}J_{m}+k_{2}J_{m}+k_{2}k_{4}+k_{1}k_{3}+k_{1}k_{4}$,
$s_{3}=-J_{m}k_{3}$, $s_{4}=-J_{m}k_{1}k_{3}$.

The plot of $p(m)$ versus $m$ is shown in figure 10 (a) (figure
10(b)) corresponding to the different $FF_{m}$ regimes shown in figure
3(a) (figure 4(a)). In figure 10(b), both the curves, the red-dotted
($J_{m}$= $1.0$) and the cyan dot-dashed ($J_{m}$= $10.0$), are
in sub-Poissonian $FF_{m}$ but with different mean value. The dot-dashed
curve for $J_{m}$= $10.0$ has higher mean but lower $FF_{m}$. The
mean mRNA level, thus, can be increased and at the same time, the
$FF_{m}$ can be decreased with reinitiation in gene transcription.

\begin{figure}[H]
\begin{centering}
\includegraphics[width=5cm,height=3cm]{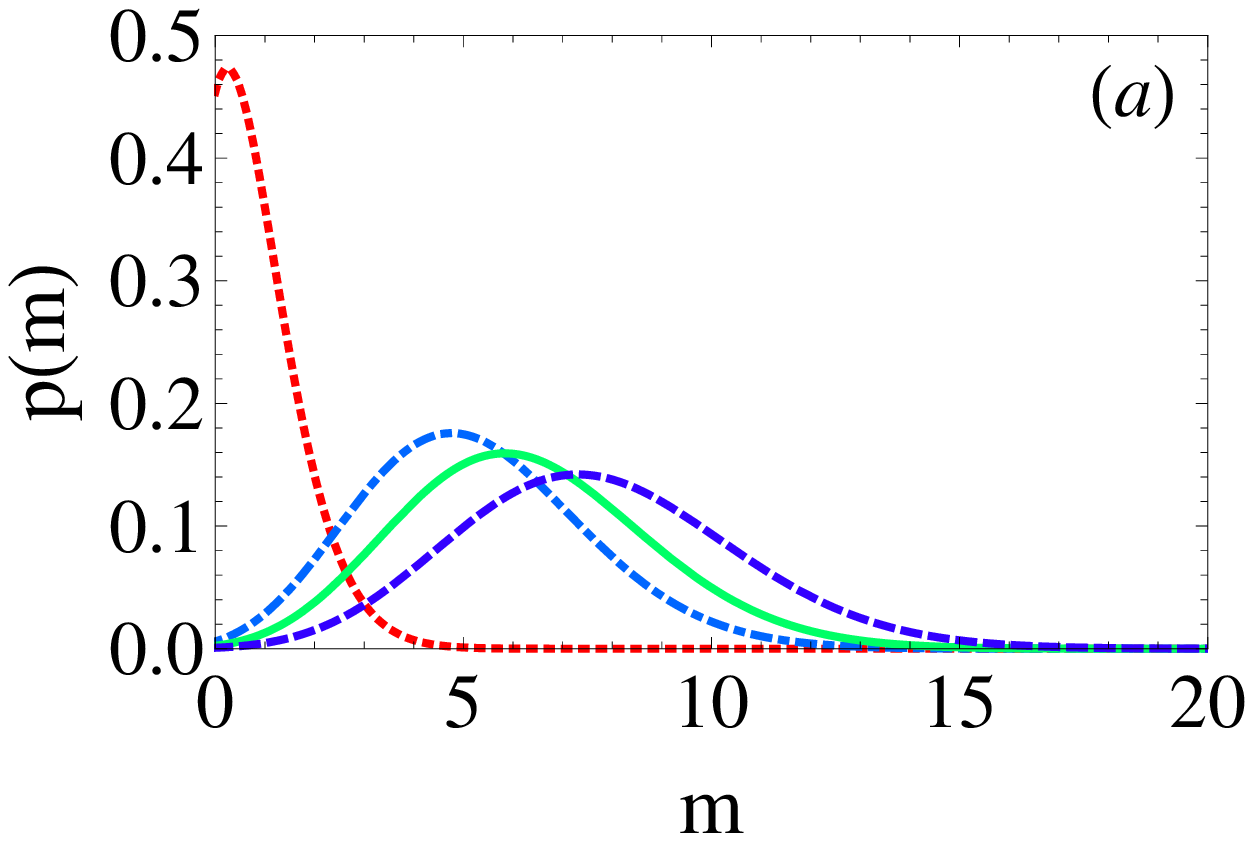}\includegraphics[width=5cm,height=3cm]{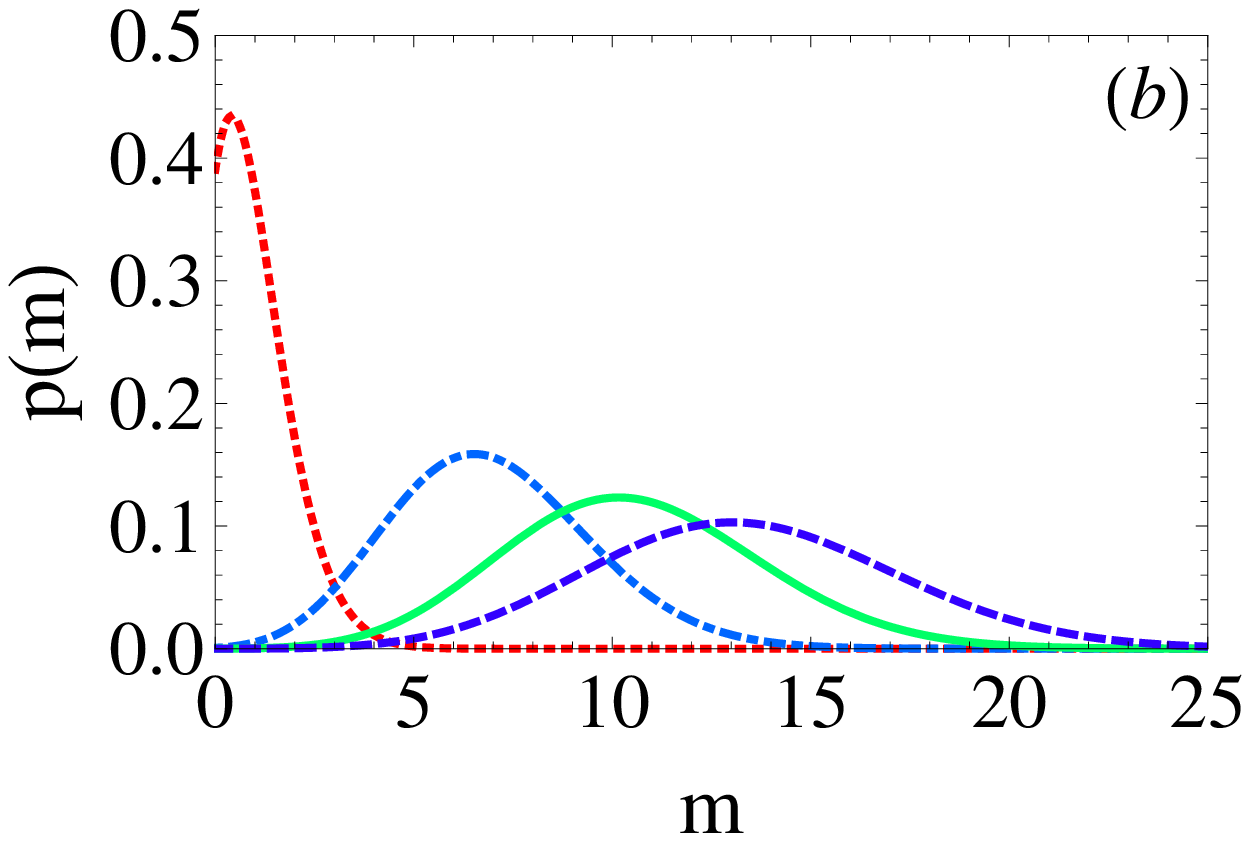} 
\par\end{centering}
Figure 10 (Color online) \textbf{(a) }Distribution of mRNA for the
rate constants as considered in figure 2, i.e., $k_{1}=6.02,$ $k_{2}=3.0125,$
$k_{3}=50.0,$ $k_{4}=10.0,$ $k_{m}=1.0$. Four different values
of $J_{m}$ are: $1.048$ (Red dotted, $<m>=0.79$, $FF_{m}=0.992$),
$8.048$ (Cyan dot-dashed, $<m>=5.22$, $FF_{m}=0.997$), $10.048$
(Green solid, $<m>=6.27$, $FF_{m}=1.00$) and $13.048$ (Blue dashed,
$<m>=7.71$, $FF_{m}=1.019$). \textbf{(b)} Distribution of mRNA level
for the rate constants: $k_{1}=10.0,\;k_{2}=10.0,\;k_{4}=1.0,\:k_{3}=50.0$.
Four different curves are for four different values of $J_{m}$: $J_{m}=1.0$
(Red dotted, $<m>=0.92$, $FF_{m}=0.971$), $J_{m}=10.0$ (Cyan dot-dashed,
$<m>=6.94$, $FF_{m}=0.913$), $J_{m}=19.0$ (Green solid, $<m>=10.55$,
$FF_{m}=0.989$) and $J_{m}=30.0$ (Blue dashed, $<m>=13.4$, $FF_{m}=1.12$). 
\end{figure}

In figure 11(a), we draw two curves by changing the GAL concentrations
only for same mean ($<m>=15.5$) but with different $FF_{m}$. The
red-dotted curve is for 10\% {[}GAL{]} with super-Poissonian $FF_{m}$
and the blue-dashed curve is for 60\% {[}GAL{]} with sub-Poissonian
$FF_{m}$. In figure 11(b), we see that mean mRNA level can be increased
without increasing the $FF_{m}$. The red-dotted curve, blue-dashed
curve and cyan-dot-dashed curves have the same $FF_{m}$ but they
are with increasing mean respectively. The green-solid curve in figure
11(b) has higher mean but lower $FF_{m}$ compared to the red-dotted
and blue-dashed curves. That kind of behaviour of mean and FF can
also be observed with lower values of GAL concentration but in narrow
range of $k_{3}$ and $J_{m}$. With higher GAL concentrations, the
$J_{m}^{c}$ becomes high and the $FF_{m}$ become sub-Poissonian
over a wide range of $k_{3}$ and $J_{m}$.

\begin{figure}[H]
\begin{centering}
\includegraphics[width=5cm,height=3cm]{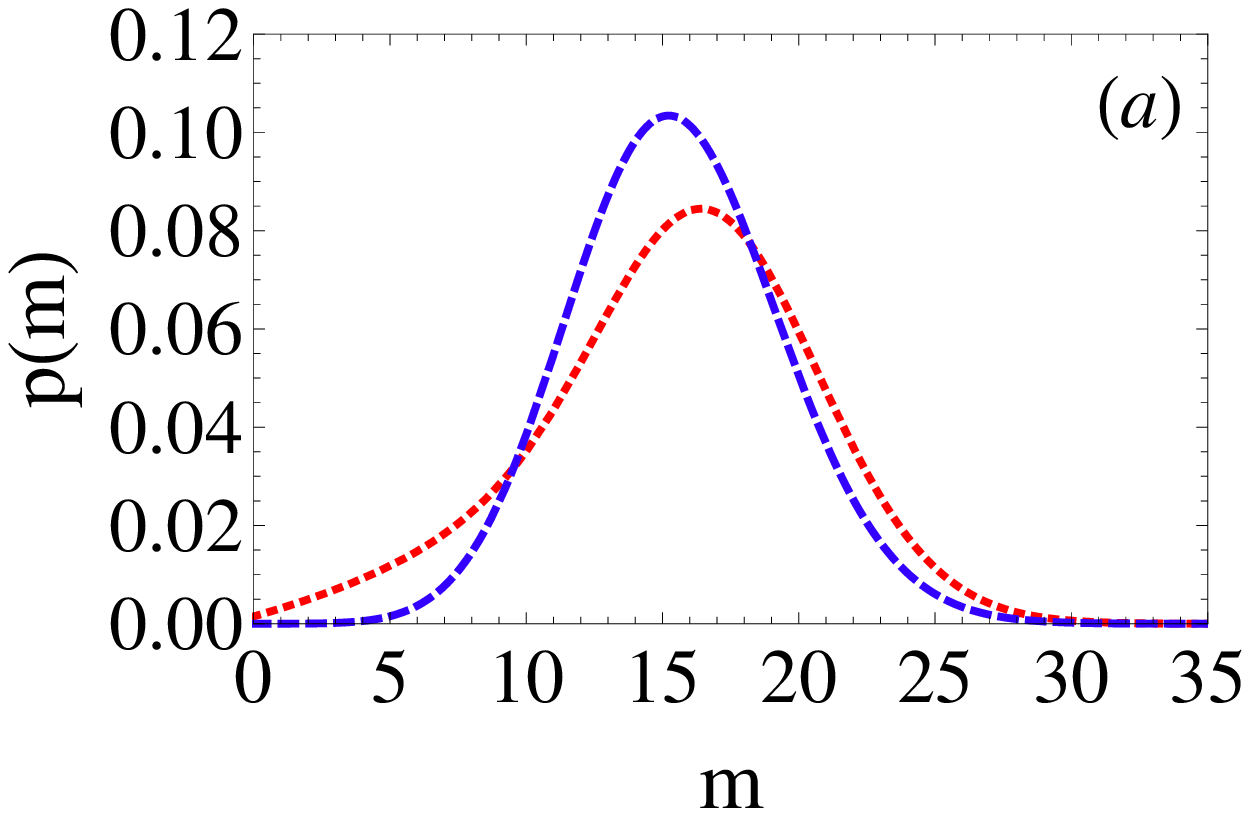} \includegraphics[width=5cm,height=3cm]{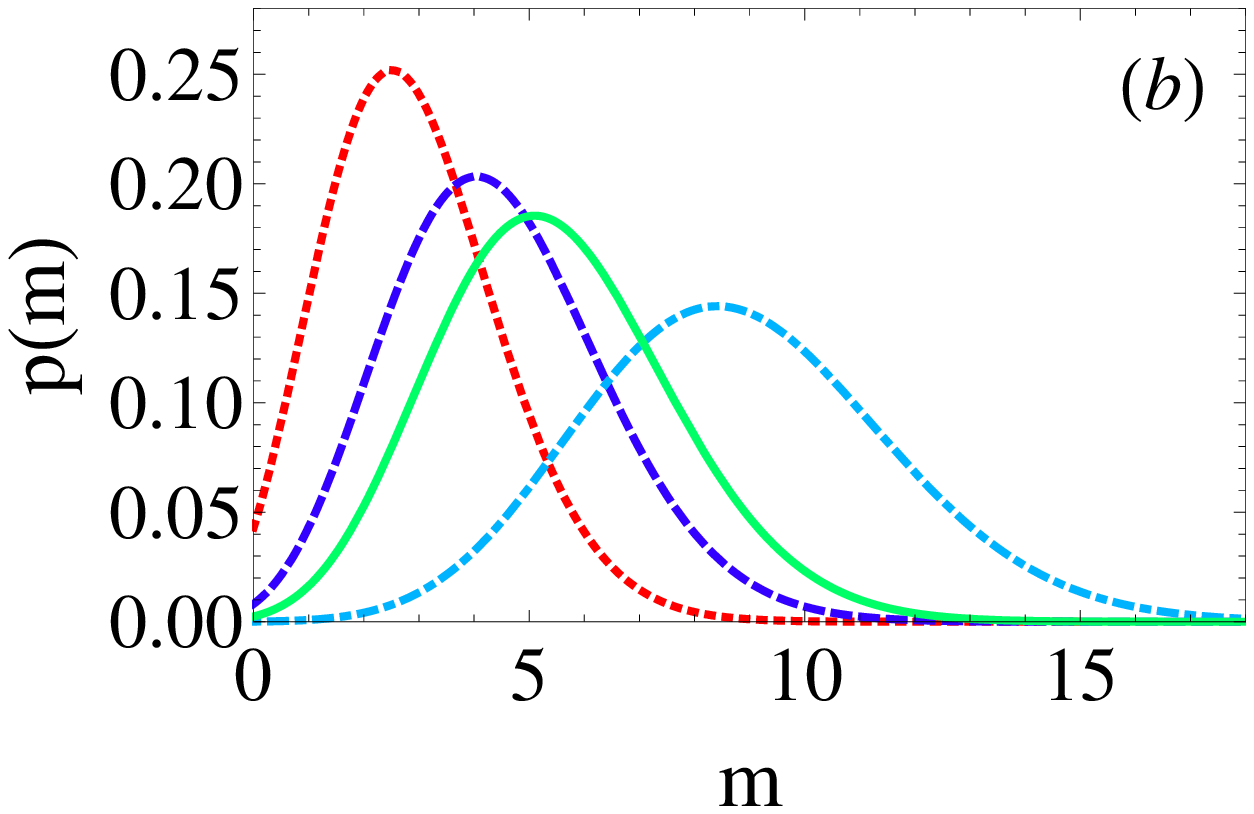} 
\par\end{centering}
Figure 11 (Color online)\textbf{ (a)} Distribution of mRNA for the
rate constants $k_{3}=50.0,$ $J_{m}=30.0,$ $k_{4}=1.0,$ $k_{m}=1.0$
with two different sets of GAL concentrations having same mean ($<m>=15.5$)
but different Fano factors. The red-dotted curve is for {[}GAL{]}
= 10\% ($FF_{m}=1.64$) and blue-dashed curve is for {[}GAL{]} = 60\%
($FF_{m}=0.95$).\textbf{ (b)} Distribution of mRNA for the {[}GAL{]}
= 60\% i.e., rate constants $k_{1}=12.02,$ $k_{2}=6.01,$ $k_{4}=1.0,$
$k_{m}=1.0$ with four different sets of $k_{3}$ and $J_{m}$: $k_{3}=20.0,$
$J_{m}=4.0$ (Red-dotted line, $<m>=2.9$, $FF_{m}=0.872$,), $k_{3}=30.0,$
$J_{m}=6.0$ (Blue-dashed line, $<m>=4.44$, $FF_{m}=0.872$) , $k_{3}=20.0,$
$J_{m}=10.0$ (Green solid line, $<m>=5.48$, $FF_{m}=0.848$) and
$k_{3}=38.0,$ $J_{m}=14.0$ (Cyan-dot-dashed line, $<m>=8.8$, $FF_{m}=0.873$) 
\end{figure}

\section{Conclusion}

Noise or stochasticity in gene expression produces fluctuations in
mRNA and protein levels. The fluctuations in protein levels can corrupt
the quality of intracellular signals, thereby, affect the cellular
functions negatively. Different studies show that the cellular system
has different mechanisms to control the inherent stochasticity. Negative
feedback is one such mechanism by which noise (FF) reduction takes
place \cite{key-119}. The network architecture, consisting of four
regulatory genes, helps to reduce the FF and CV during aging \cite{key-122F}.
A coherent feed forward network with three genes can also show least
variance level in proteins \cite{key-126}. In this study, we observed
that the reinitiation of transcription by RNAP can also reduce the
FF during gene transcription.

To study the contribution of reinitiation in the variability of mRNA
and protein levels, we considered a simple two-state model with reinitiation
of RNAP (figure 1) and calculated the Fano factor in mRNA and protein
levels using simple mathematics. Our model network is similar to the
regulatory network of GAL1{*} yeast promoter with full ATC induction
\cite{key-111}. So, we compared our analytically calculated FF of
protein levels with the experimental results of Blake et al. at full
ATC induction and found excellent agreement. Then we analyzed the
Fano factor in mRNA levels and observed three different phases viz.,
Poissonian, sub-Poissonian and super-Poissonian phases'. We found
that there exists a critical value $J_{m}^{c}$ for which the $FF_{m}$
is always Poissonian. For $J_{m}<J_{m}^{c}$ ($J_{m}>J_{m}^{c}$)
the $FF_{m}$ lies in the sub-Poissonian (super-Poissonian) phase.
That result is distinct from the well-studied constitutive and two-state
gene activation model. In those two models, the FF in mRNA levels
is Poissonian and super-Poissonian respectively \cite{key-121,key-124B}.
The minimum value of $FF_{m}$ in two-state model can be very close
to unity but never less than unity.

In our model network, we included the reinitiation step into the two-state
gene activation model and found sub-Poissonian Fano factor regime
in mRNA levels. The reinitiation process, therefore, not only removed
the extra additive term in $FF_{m}$ due to random transitions between
the gene states in two-state model completely rather decreased it
further below unity. That shows that the reinitiation step in gene
expression has the strong ability to control the fluctuations in mRNA
and protein levels. In the constitutive and two-state model, the mRNA
synthesis takes place with constant rate from the active state of
the gene. But, the reinitiation process in gene transcription adds
a refractory period after each mRNA synthesis and that reduces the
effective rate of mRNA synthesis and variance though the mean mRNA
level increases. This results in the\textbf{\textcolor{black}{{} }}$FF_{m}$
going below unity in gene transcription with reinitiation.

In the two-state gene activation model, the activation ($k_{1}$)
and deactivation ($k_{2}$) rate constants mostly regulate the FF
in mRNA level. The fast (slow) transition between the active and inactive
states decreases (increases) the FF in mRNA level \cite{key-112}.
Thus, sub-Poissonian $FF_{m}$ in our model network is favorable with
higher values of $k_{1}$ and $k_{2}$ (figure 6(a)). It can be shown
that the mean mRNA level decreases (increases) with the increase of
$k_{2}$ ($k_{1}$). But with high $k_{1}$, the $k_{2}$ has the
least effect on mean mRNA levels. In the experiment of Blake et al.,
$k_{1}$ and $k_{2}$ are both functions of GAL concentrations and
can be made high. The high $k_{1}$ and $k_{2}$ leads to mean mRNA
level going up to almost saturation but the $FF_{m}$ decreasing to
the sub-Poissonian regime (figures 9(a) and 9(b)). We can visualize
that result from the distributions of mRNA levels in figure 11(a).
Therefore, one can adjust the $FF_{m}$ to sub-Poissonian regime without
changing the mean mRNA by regulating the GAL concentrations in the
system.

In our model network, the desired mean and $FF_{m}$ can also be obtained
by choosing the rate constants $k_{3}$ and $J_{m}$ properly (figure
8(c), figure 5(a) and figure 11(b)). The $FF_{m}$ shows a dip when
plotted with $k_{3}$ and $J_{m}$ whereas the mean increases with
those rate constants (figure 5(a) and 8(c)). Therefore, mean can be
increased keeping $FF_{m}$ fixed (the red-dotted line, the blue-dashed
line and cyan-dot-dashed line) or mean can be increased while $FF_{m}$
decreases (red-dotted line, the green solid line) (figure 11(b)).
Like $k_{1}$ and $k_{2}$ in the experiment of Blake et al., the
rate constants $k_{3}$ and $J_{m}$ are not controlled externally.
But cell can change their values to fulfill its requirements. There
are several instances that the cellular system has the ability to
adapt to the unpredictable environmental changes by regulating its
internal states, mean protein levels, the fluctuations about the mean
levels etc. \cite{key-122A,key-122B,key-122D,key-122E,key-122F,key-127,key-128}.
This ability is crucial for survival and proper functioning of cells.


\begin{thebibliography}{10}
\bibitem{key-109}Elowitz M B\emph{, }Levine A J, Siggia E D and Swain
P S 2002 Stochastic gene expression in a single cell \emph{Science}
\textbf{297,} 1183-1186.

\bibitem{key-110}Ozbudak E M, Thattai M, Kurtser I, Grossman A D
and Oudenaarden A van 2002 Regulation of noise in the expression of
single gene \emph{Nature Genet.} \textbf{31}, 69-73.

\bibitem{key-110A}Swain P S, Elowitz M B and Siggia E D 2002 Intrinsic
and extrinsic contributions to stochasticity in gene expression \emph{PNAS}
\textbf{99}, 12795-12800.

\bibitem{key-111}Blake W J, Kaern M, Cantor C R and Collins J J 2003
Noise in eukaryotic gene expression \emph{Nature} \textbf{422}, 633-637.

\bibitem{key-111A}Blake W J, Bal$\acute{a}$zsi G, Kohanski M A,
Isaas F J, Murphy K F, Kuang Y, Cantor C R, Walt D R and Collins J
J 2006 Phenotypic consequences of promoter-mediated transcriptional
noise \emph{Molecular Cell} \textbf{24}, 853-865.

\bibitem{key-112}Karmakar R and Bose I 2004 Graded and binary responses
in stochastic gene expression \emph{Phys. Biol.}\textbf{ 1,} 197-204.

\bibitem{key-113}Raser J M and O'Shea E K 2005 Noise in gene expression:
origins, consequences, and control \emph{Science} \textbf{309,} 2010-2013.

\bibitem{key-114}Hornos J E M, Scshultz D, Innocentini G C P, Wang
J, Walczak A M, Onuchic J N and Wolynes P G 2005 Self regulating gene:
An exact solution \emph{Phys. Rev. E} \textbf{72}, 051907/1-5.

\bibitem{key-115}Kaern M, Elston T C, Blake W J and Collins J J 2005
Stochasticity in gene expression: from theories to phenotypes \emph{Nat.
Rev. Genet}.\textbf{ 6,} 451-464.

\bibitem{key-116}Raj A, Peskin C S, Tranchina D, Vargas D Y and Tyagi
S 2006 Stochastic mRNA synthesis in mammalian cells \emph{PLoS Biol.}
\textbf{4}, e309/1707-1719.

\bibitem{key-117}Shahrezaei V and Swain P S 2008 Analytical didtributions
for stochastic gene expression \emph{PNAS} \textbf{105,} 17256-17261.

\bibitem{key-117B}Rossi F M V, Kringstein A M, Spicher A, Guicherit
O M and Blau H M 2000 Transcriptional Control: Rheostat Converted
to On/Off Switch \emph{Molecular Cell} \textbf{6}, 723--728.

\bibitem{key-118}Karmakar R 2010 Conversion of graded to binary responses
in an activator-repressor system \emph{Phys. Rev. E}. \textbf{81}
021905-12.

\bibitem{key-119}Kumar N, Platini T and Kulkarni R V 2014 Exact distribution
for stochastic gene expression models with bursting and feedback \emph{Phys.
Rev. Lett.} \textbf{113}, 268105/1-5.

\bibitem{key-121}Paulsson J 2005 Models of stochastic gene expression
\emph{Physics of Life Review} \textbf{2}, 157-175.

\bibitem{key-121A}Sanchez A, Choubey S and Kondev J 2013 Regulation
of Noise in Gene Expression \emph{Annu. Rev. Biophys.} \textbf{42},
469--491.

\bibitem{key-122A}Maamar H, Raj A and Dubnau D 2007 Noise in Gene
Expression Determines Cell Fate in Bacillus subtilis \emph{Science}
\textbf{317}, 526-529.

\bibitem{key-122B}Acar M, Mettetal J T and Oudenaarden A van 2008
Stochastic switching as a survival strategy in fluctuating environments
\emph{Nat. Genet.} \textbf{40}, 471-475.

\bibitem{key-122C}Magee J A, Abdulkadir S A and Milbrandt J 2003
Haploinsufficiency at the Nkx 3.1 locus. A paradigm for stochastic,
dose-sensitive gene regulation during tumour initiation \emph{Cancer
Cell}, \textbf{3}, 273--283.

\bibitem{key-122D}Weinberger L S, Burnett J C, Toettcher J E, Arkin
A P and Schaffer D V 2005 Stochastic Gene Expression in a Lentiviral
Positive-Feedback Loop: HIV-1 Tat Fluctuations Drive Phenotypic Diversity
\emph{Cell}, \textbf{122}, 169--182.

\bibitem{key-122E}Fraser H B, Hirsh A E, Giaever G, Kumm J and Eisen
M B 2004 Noise minimization in eukaryotic gene expression \emph{PLoS
Biology }\textbf{2}, 1-5.

\bibitem{key-122F}Liu P, Song R, Elison G L, Peng W and Acar M 2017
Noise reduction as an emergent property of single-cell aging \emph{Nature
Communications} \textbf{8}, 1-13.

\bibitem{key-13}Bundschuh R, Hayot F and Jayaprakash C The Role of
Dimerization in Noise Reduction of Simple Genetic Networks \emph{J.
Theo. Biol.} \textbf{220}, 261-269.

\bibitem{key-14}Swain P S Efficient Attenuation of Stochasticity
in Gene Expression Through Post-transcriptional Control \emph{J. Mol.
Biol. }\textbf{344}, 965-976.

\bibitem{key-3}Alberts B, Johnson A, Lewis J, Raff M, Roberts K and
Walters P 2002 \emph{Molecular Biology of the Cell}, Garland Science,
UK.

\bibitem{key-6}Barberis A and Petrascheck M 2003 Transcription activation
in eukaryotic cells \emph{Encyclopedia of life sciences; }doi:10.1038/npg.els.0003303

\bibitem{key-123}Struhl K 1999 Fundamentally different logic of gene
regulation in eukaryotes and prokaryotes \emph{Cell} \textbf{98},
1-4.

\bibitem{key-124}Liu B, Yuan Z, Aihara K and Chen L 2014 Reinitiation
enhances reliable transcriptional responses in eukaryotes \emph{J.
R. Soc. Interface} \textbf{11}, 0326/1-11.

\bibitem{key-124A}Sanchez A and Kondev J 2008 Transcriptional control
of noise in gene expression \emph{PNAS} \textbf{105,} 5081-5086.

\bibitem{key-124B}Jones D L, Brewster R C and Phillips R 2014 Promoter
architecture dictates cell-to-cell variability in gene expression
\emph{Science} \textbf{346}, 1533-1536.

\bibitem{key-124C}Munsky B, Neuert G, Oudenaarden A van 2012 Using
gene expression noise to understand gene regulation \emph{Science}
\textbf{336}, 183-187.

\bibitem{key-125}van Kampen N G 1985 \emph{Stochastic Processes in
Physics and Chemistry}, North-Holland, Amsterdam.

\bibitem{key-125A}Gillespie D T 1977 Exact stochastic simulation
of Coupled Chemical Reactions \emph{J. Phys. Chem.} \textbf{81}, 2340-2361.

\bibitem{key-126}Ghosh B, Karmakar R and Bose I 2005 Noise characteristics
of feed forward loops \emph{Phys. Biol.} \textbf{2}, 36--45.

\bibitem{key-127}Halpern K B, Tanami S, Landen S, Chapal M, Szlak
L, Hutzler A, Nizhberg A, and Itzkovitz S 2015 Bursty Gene Expression
in the Intact Mammalian Liver \emph{Mol. Cell} \textbf{58}, 1--10.

\bibitem{key-128}Briat C, Gupta A, Khammash M. 2016 Antithetic integral
feedback ensures robust perfect adaptation in noisy biomolecular networks
\emph{Cell Syst}. \textbf{2}, 17--28. 
\end{thebibliography}
\end{document}